\documentclass[aps,prl,twocolumn,superscriptaddress,10pt,article,showpacs]{revtex4-2}

\usepackage{blindtext}
\usepackage{lipsum}
\usepackage{graphics}
\usepackage{amsmath}
\usepackage{graphicx}
\usepackage{graphics}
\usepackage{amssymb}
\usepackage{verbatim}
\usepackage{physics}
\usepackage{float}
\usepackage[normalem]{ulem}
\usepackage[dvipsnames]{xcolor}
\usepackage{bbm}
\usepackage{bm}
\usepackage{enumitem} 
\usepackage{booktabs}
\usepackage{tikz}
\usetikzlibrary{positioning, decorations.pathreplacing, calligraphy, fit, backgrounds, arrows.meta, calc}

\definecolor{blueviolet}{rgb}{0.2, 0.2, 0.6}
\definecolor{webgreen}{rgb}{0,.5,0}
\definecolor{webbrown}{rgb}{.6,0,0}
\usepackage[bookmarks=false,
	colorlinks=true, 
	urlcolor=webbrown, 
	linkcolor=blueviolet, 
	citecolor=webgreen,
	pdfstartpage=1,
	pdfstartview={FitH},  
	bookmarksopen=false
	]{hyperref}

\makeatletter 
    
\renewcommand\onecolumngrid{
\do@columngrid{one}{\@ne}%
\def\set@footnotewidth{\onecolumngrid}
\def\footnoterule{\kern-6pt\hrule width 1.5in\kern6pt}%
}

\newcommand{\be}{\begin{equation}}
\newcommand{\ee}{\end{equation}}
\def\bea{\begin{eqnarray}}
\def\eea{\end{eqnarray}}

\def\ba{\begin{array}}
\def\ea{\end{array}}
\def\Tr{\mathrm{Tr}}

\begin{document}

\title{Preparing High-Fidelity Thermofield Double States}

\author{Brian J. J. Khor}
\email{bkhor@brandeis.edu}
\affiliation{Martin Fisher School of Physics, Brandeis University, Waltham, Massachusetts 02453, USA}

\author{Nadie LiTenn}
\email{nadie@brandeis.edu}
\affiliation{Martin Fisher School of Physics, Brandeis University, Waltham, Massachusetts 02453, USA}

\author{Martin Sasieta}
\email{msasieta@berkeley.edu}
\affiliation{Leinweber Institute for Theoretical Physics, University of California, Berkeley, California 94720, USA}

\author{Brian Swingle}
\email{bswingle@brandeis.edu}
\affiliation{Martin Fisher School of Physics, Brandeis University, Waltham, Massachusetts 02453, USA}

\begin{abstract}

A major promise of quantum computers is the controlled preparation of many-body quantum states beyond the reach of efficient classical computation. Among the most important targets are thermal mixed states and their thermofield double (TFD) purifications, which play central roles in quantum many-body physics and quantum gravity. For target systems with a bounded energy spectrum that obey the eigenstate thermalization hypothesis (ETH), we present a parent Hamiltonian built from two copies of the target Hamiltonian and ultra-local couplings between the copies, which we argue is gapped with a ground state that approximates a TFD state of the target Hamiltonian. By adiabatically evolving down from strong coupling, we can thus prepare a high-fidelity TFD state. We study two variants of the parent Hamiltonian using numerical methods in two classes of models: mixed field Ising models in one and two dimensions and non-local ``spin Sachdev-Ye-Kitaev'' models. In the simpler variant, the parent Hamiltonian ground state has high overlap with a TFD for system sizes accessible to near-term quantum devices. However, the global overlap decays exponentially with the number of qubits, with a small error per degree of freedom. The second variant introduces an additional penalty term which can be tuned to reduce or remove the decay of the overlap with system size. Together with a general ETH-based analysis, these results suggest a broadly applicable method for TFD preparation that is not limited to particular temperatures or geometric locality. 

\end{abstract}

\maketitle

\section{Introduction}

Many physical systems can be idealized as existing in thermal equilibrium at some temperature. To simulate the physical properties of such systems on a quantum computer, it is useful to be able to prepare such thermal states efficiently \cite{Terhal2000b}. A simple idea is to imitate Nature by coupling the system of interest to a heat bath, but this may not be the best approach. For example, modeling the bath may require many additional qubits or otherwise involve substantial overhead \cite{Temme2011,Poulin2009, Chen:2023}. There are also cases where the natural bath dynamics leads to slow cooling, as in the phenomenon of critical slowing down near a second order phase transition \cite{Hohenberg1977}. In such cases, artificial dynamics, such as Swendsen–Wang dynamics \cite{Swendsen1987}, may be much more efficient in thermalizing the system. Additionally, for a number of applications, one wants not just the mixed thermal state but a particular minimal purification of it known as a thermofield double (TFD) state \cite{Maldacena:2001kr, maldaqi, brown_quantum_2023, Gao_2017, Wu:2019hnq, Martyn:2018wli, Chapman2019a}. Such TFD states are not something that Nature automatically provides. Hence, it is broadly important to find and study efficient methods to characterize and prepare thermal mixed states and TFD states on programmable quantum devices.

In this paper, we propose a method to prepare TFD states that is plausibly efficient for any system that (i) has a bounded energy spectrum, (ii) obeys the eigenstate thermalization hypothesis (ETH) \cite{PhysRevA.43.2046,Srednicki:1994mfb}, and (iii) does not exhibit a thermal phase transition at or above the temperature of interest. At the core of our proposal is a simple \emph{parent Hamiltonian} built from the Hamiltonian of interest, with the empirical property that its ground state has significant overlap with the TFD state at a tunable temperature. By studying this parent Hamiltonian across a range of models, including 1- and 2-dimensional systems and systems with all-to-all interactions, we offer strong numerical and analytical evidence for the wide applicability and usefulness of our proposal.

\begin{figure}[b]
\centering
\vspace{-.2cm}
\includegraphics[width = .48\textwidth]{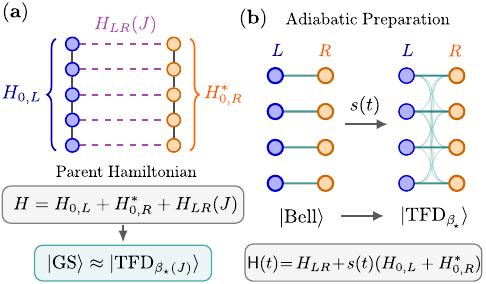} 
\caption{Parent Hamiltonian construction and adiabatic preparation for a spin chain.
(a) Coupling the $L$ and $R$ systems with strength $J$ yields a ground state approximating a finite temperature TFD state at an inverse temperature $\beta_\star(J)$.
(b) Starting at strong coupling, where the ground state is the infinite temperature TFD, $|\rm Bell\rangle$ (a collection of Bell pairs), an adiabatic ramp prepares the finite temperature TFD state.}
\label{fig:parent-setup}
\end{figure}

The setup is illustrated in Fig.~\ref{fig:parent-setup}. We start from a target Hamiltonian $H_0$ whose TFD state we are interested in, 
\be
    |{\rm TFD}_\beta\rangle
    =
    \frac{1}{\sqrt{Z(\beta)}}
    \sum_n e^{-\beta \epsilon_n/2}\,
    |\epsilon_n\rangle |\epsilon_n^*\rangle \,,
\ee
where $\beta$ is the inverse temperature, $|\epsilon_n\rangle$ is an energy eigenstate of $H_0$, $|\epsilon_n^*\rangle=\Theta|\epsilon_n\rangle$ is the image of $|\epsilon_n\rangle$ under an antiunitary map $\Theta$, and  $Z(\beta)=\Tr\,e^{-\beta H_0}$ is the canonical partition function. The parent Hamiltonian is defined on two copies, left ($L$) and right ($R$), of the original system and features a special $LR$ coupling, see Fig.~\ref{fig:parent-setup} $(\rm a)$ and Eq. \eqref{eq:Hamiltonianprop}. We then demonstrate that the ground state $|\rm GS\rangle$ of the parent Hamiltonian has significant overlap with a TFD state of $H_0$ at a temperature controlled by the strength of the $LR$ coupling.

Moreover, we provide numerical and analytical evidence that the coupled Hamiltonian has a gap that persists to the strong coupling (infinite temperature) limit, provided $H_0$ does not exhibit any thermal phase transitions in the relevant temperature range and obeys the eigenstate thermalization hypothesis. The existence of a robust gap then yields a method to prepare the TFD state of interest by adiabatic evolution of the coupled Hamiltonian from strong coupling to the coupling (and hence $\beta$) of interest, as represented in Fig. \ref{fig:parent-setup} $(\rm b)$. These results explain our conjecture that the criteria (i-iii) are sufficient for the protocol to be efficient.

A central challenge for these methods is to scale to large system sizes while maintaining high fidelity between the ground state and the target TFD state. In the basic construction, the $LR$ coupling is ultra-local between the two copies of the system. We find that the fidelity between the ground state and the target TFD state decays exponentially with system size.  To address this, we introduce a modified version of the parent Hamiltonian by adding an extra penalty term. The modified ground state yields a much closer approximation to the TFD state at all temperatures, in the sense that the fidelity remains large. Although this modification of the parent Hamiltonian introduces bi-local couplings, we argue that the resulting adiabatic preparation protocol remains efficient, in the sense that it can still be implemented with a circuit whose depth is polynomial in the system size, $N$.

Our work comes in the context of a growing body of work on thermal-state and TFD preparation, including variational and optimization-based schemes \cite{Martyn:2018wli,Wu:2019hnq,Verdon2019,Sagastizabal2021,sewell2022thermalmultiscaleentanglementrenormalization,Araz:2024xkw,Zhu:2020gro}, imaginary-time methods \cite{Motta:2019qite,Sun:2021prxqfiniteT}, quantum Gibbs-sampling algorithms \cite{Poulin2009,Temme2011,Chen:2023}, and dissipative or engineered-bath protocols \cite{Chen:2023zpu,Rouze:2025,Chen:2025fax,Ding2025,Rall2023,lloyd2026quantumthermalstatepreparation}. As noted above, our method can be classified as a parent Hamiltonian approach \cite{maldaqi,Cottrell_2019,Alet:2020ehp,Su:2020zgc,Failde:2023bho,Bentsen:2023xlu,schuster2025cooling}, which goes back at least to \cite{feiguin2013hermitiannonhermitianthermalhamiltonians} and \cite{Swingle_2016_mixed_s} and is particularly well motivated by developments in holographic models of quantum gravity. There, the TFD state plays a distinguished role as the microscopic description of a gravitational ``wormhole'' connecting the ``left'' ($L$) and ``right'' ($R$) asymptotic regions of space \cite{Maldacena:2001kr} and serves as the natural resource for teleportation protocols that probe wormhole traversability \cite{Gao_2017,Maldacena:2017axo,maldaqi,brown_quantum_2023,nezami_quantum_2023,Lu:2026dns}.

We begin in Section \ref{sec:parent} by introducing the parent Hamiltonian and discussing its basic properties, including the quantities that will organize our analysis: the optimal TFD fidelity, the associated inverse temperature, and the gap. We also summarize the different kinds of evidence developed in the rest of the paper, combining tensor network numerics, exact diagonalization, and analytic arguments.

In Section \ref{sec:ising} we study the parent Hamiltonian in detail for the mixed field Ising chain, using matrix product state methods \cite{Schollwoeck2011} to reach system sizes up to $N\lesssim 100$. For moderate system sizes, $N\sim 30$, we find that the ground state has substantial overlap with a TFD state, with fidelity exceeding $0.6$ across all $LR$ couplings we consider. In Section \ref{sec:other} we complement this case study with exact diagonalization results in additional models: the two-dimensional mixed field Ising model and a spin Sachdev-Ye-Kitaev (SYK) model. The results demonstrate that a high overlap with the TFD also appears at small system sizes in both local and nonlocal settings, and across different dimensions. In Section \ref{sec:random_matrix_main} we present analytic results for a random matrix version of the parent Hamiltonian.

Having established these basic properties, we turn in Section \ref{sec:adiabatic} to the corresponding adiabatic state-preparation protocol. Focusing again on the mixed field Ising model, we study a simple linear ramp and numerically find that, in the long-time limit, the adiabatically prepared state approaches the true ground state with large fidelity. We also show that more modest circuit depths already suffice to prepare a TFD-like state with $O(1)$ fidelity, making the protocol promising for near-term implementations.

In Section \ref{sec:scaling-problem} we investigate the behavior at larger system sizes and show that the overlap with the TFD decays exponentially with $N$, although with a numerically small exponent. This reveals a genuine scalability problem at sufficiently large $N$, and also exposes a limitation of the general argument of \cite{Cottrell_2019}, which we trace to the leakage of the ground state wavefunction out of the diagonal-in-energy subspace of the doubled Hilbert space. 

In Section \ref{sec:penalty-term} we address this problem by introducing an improved parent Hamiltonian with an energy-mismatch penalty term that suppresses this leakage and substantially enhances the overlap with the TFD. We show that, for sufficiently large penalty strength, the fidelity can remain nearly constant as $N$ increases. In Section \ref{sec:syk_main} we then apply the same idea to the Maldacena-Qi model and show how the penalty term improves the TFD fidelity of the corresponding wormhole ground state.

We conclude in Section \ref{sec:outlook} with a discussion of open questions and possible future directions.

Three appendices develop the analytic picture in greater detail. Appendix \ref{sec:theory} revisits the general argument for the parent Hamiltonian in chaotic systems and clarifies the origin of the finite-$K$ fidelity loss. Appendix \ref{sec:rmt} analyzes a random matrix version of the construction, which provides a useful solvable comparison. Appendix \ref{sec:SYK} studies the effect of the penalty term in the Maldacena-Qi model.

\section{Coupled parent Hamiltonian}
\label{sec:parent}

Given a quantum many-body system with Hamiltonian $H_0$, we consider two copies of the system, denoted left ($L$) and right ($R$), and couple them through a collection of $K$ Hermitian operators $\{\mathcal{O}_1,...,\mathcal{O}_K\}$. This defines the Hamiltonian
\be\label{eq:Hamiltonianprop}
H = H_{0,L} + H^*_{0,R} + \dfrac{JN}{2K} \sum_{\alpha=1}^K\left(\mathcal{O}_{\alpha,L} -\mathcal{O}_{\alpha,R}^* \right)^2\,,
\ee
where the star denotes conjugation by the antiunitary map, $\mathcal{O}_R^* = \Theta \,\mathcal{O}_L\, \Theta^{-1}$. The parameter $N$ is an extensive measure of system size. We normalize the operators so that their microcanonical variance is non-extensive, which ensures that each term in $H$ is itself extensive. For simplicity, we take all operator couplings to be equal to $J$, as is natural in homogeneous systems, although the construction can be extended straightforwardly to nonuniform couplings.

The basic proposal is that the ground state $|{\rm GS}\rangle$ of the coupled Hamiltonian \eqref{eq:Hamiltonianprop} approximates a TFD state $\ket{{\rm TFD}_{\beta_\star}}$ at a coupling-dependent inverse temperature $\beta_\star = \beta_\star(J)$, and that the Hamiltonian \eqref{eq:Hamiltonianprop} remains gapped in the thermodynamic limit.

There are two simple limits in which the structure of the ground state is transparent. At zero $LR$ coupling, $J=0$, the ground state of \eqref{eq:Hamiltonianprop} is the ``zero-temperature TFD'',
\be
\lim\limits_{J\to 0}|{\rm GS}\rangle = |0\rangle_L|0^*\rangle_R\,,
\ee
where $|0\rangle$ is the ground state of $H_0$. Thus $\beta_\star(0)=\infty$. In this limit, the gap of $H$ is simply the gap of $H_0$.

At the opposite extreme, in lattice systems with infinite $LR$ coupling, $J\to\infty$, the ground state is the ``infinite temperature TFD'',
\be
\lim\limits_{J\to\infty}|{\rm GS}\rangle =\ket{\infty}\,=\,\dfrac{1}{\sqrt{Z(0)}}\sum_n |\epsilon_n\rangle |\epsilon_n^*\rangle\,,
\ee
which satisfies $\left(\mathcal{O}_{\alpha,L}-\mathcal{O}_{\alpha,R}^*\right)\ket{\infty}=0$ for all $\alpha=1,\dots,K$. This state is the unique ground state provided the operators $\{\mathcal{O}_\alpha\}$ do not all commute with one another and do not preserve a common symmetry. Therefore, $\beta_\star(\infty)=0$. Moreover, studies of the interaction term in specific systems \cite{Jian:2022pvj,Guo:2024zmr} indicate that the parent Hamiltonian remains gapped in the thermodynamic limit, with the size of the gap controlled mainly by the degree of locality of the operators $\mathcal{O}_\alpha$. We make a particular choice of the $\mathcal{O}$s below and comment further on this choice in the Outlook \ref{sec:outlook}.

At finite $LR$ coupling $J$, the heuristic picture is simple. The terms $H_{0,L}+H_{0,R}^*$ favor low-energy states, while the operator-dependent interaction penalizes states with misaligned left-right correlations. The competition between these two effects is therefore expected to select a state close to a TFD at a finite, coupling-dependent temperature, provided no thermal phase transition intervenes. 

To test how well this picture works, we define the maximal TFD fidelity
\be\label{eq:fidelity}
\mathcal{F}:= \max_{\beta}\,|\!\bra{\rm TFD_\beta}\ket{\rm GS}\!|^2 \,,
\ee
the corresponding optimal inverse temperature
\be\label{eq:betastar}
\beta_\star(J):=\arg\max_\beta \, \big|\!\bra{{\rm TFD}_\beta}\ket{{\rm GS}}\!\big|^2 \,,
\ee
and the energy gap
\be\label{eq:gapprop}
\Delta := E_1-E_0\,,
\ee
where $E_0$ and $E_1$ are the two lowest eigenvalues of $H$. These are the main quantities we study analytically and numerically throughout this paper.

It is possible that the map $\beta_\star(J)$ is not smooth, for example, it may have jumps, although we expect it to always be smooth at sufficiently large $J$ and small $\beta$. If the map is not smooth, then it may only be possible to prepare TFD states by adiabatic continuation up to some maximum $\beta$. We find that the map is smooth at all accessible $\beta$ in the cases we considered. However, in the Outlook \ref{sec:outlook} we speculate on some cases where the map may fail to be smooth.

Later in Section \ref{sec:penalty-term} we will introduce an additional penalty term into Eq. \eqref{eq:Hamiltonianprop} which takes the form
\begin{equation}
    \frac{\xi}{2N} (H_{0,L} - H_{0,R}^*)^2.
    \label{eq:penalty_term}
\end{equation}
By taking $\xi$ large, this penalty term can improve the overlap between the parent Hamiltonian ground state and a TFD state.

\paragraph{Previous proposals ---} Cottrell et al. \cite{Cottrell_2019} argued generally that for chaotic $H_0$ satisfying ETH, the ground state of the Hamiltonian \eqref{eq:Hamiltonianprop} has finite fidelity with the TFD in the thermodynamic limit $N\to \infty$,
\be\label{eq:fidelityprop}
\mathcal{F}\;\stackrel{?}{\sim}\; O(N^0)\,,
\ee
so that the error per degree of freedom in approximating $\ket{{\rm TFD}_{\beta_\star}}$ by $\ket{\rm GS}$ vanishes. For critical systems, \cite{Cottrell_2019} further argued that the gap scales as $\Delta \sim 1/\beta_\star$.

In Appendix \ref{sec:theory}, we revisit this general argument, clarify the assumptions under which it holds, and point out its limitations. In particular, while \cite{Cottrell_2019} suggested that small $K$ may be sufficient, we argue instead that the mechanism requires $K$ to be large and to scale with $N$. Even then, as we will show throughout this paper, a more careful analysis indicates that \eqref{eq:fidelityprop} is not satisfied.

Maldacena and Qi \cite{maldaqi} constructed a specific example of this parent Hamiltonian for two SYK systems coupled by a bilinear fermion interaction, with $K=N$, motivated by its gravitational interpretation (for details, see Appendix \ref{sec:SYK}). At sufficiently small coupling, such that the associated temperature lies in the nearly conformal phase of SYK, \cite{maldaqi} provided a semiclassical wormhole description of the ground state that resembles the TFD. In addition, for $4$-body SYK, the gap scales as $\Delta \sim J^{2/3}$  and $\beta_\star \sim 1/\Delta$ (in units of the SYK coupling strength).

Moreover, \cite{maldaqi} provided a gravity calculation showing that as $N\to \infty$, the fidelity behaves as
\be\label{eq:fidelityMQo}
\mathcal{F}\;{\sim}\; e^{-a N}\,,
\ee
for some coefficient $a$ that depends on the $LR$ coupling and other details of the model. We review this calculation in Appendix \ref{sec:SYK}.

Eq. \eqref{eq:fidelityMQo} indicates a constant error per degree of freedom in approximating the TFD by the ground state $|\rm GS\rangle$. However, in this model, the coefficient $a$ is small (see Appendix \ref{sec:SYK}), so the resulting loss of fidelity is not readily visible at the moderately small system sizes usually accessible when doing finite $N$ numerics. In the next section, we present numerical and analytical evidence suggesting that, for the small system sizes accessible to near-term quantum devices, this fidelity loss is likewise not a significant obstacle in other systems. For previous numerical studies at small system sizes in SYK and other models, see also \cite{maldaqi,Cottrell_2019,Garcia-Garcia:2019poj,Lantagne-Hurtubise:2019svg,Alet:2020ehp,Su:2020zgc,Failde:2023bho,schuster2025cooling}.

However, the exponential decay of the fidelity is clearly incompatible with the general proposal \eqref{eq:fidelityprop}. In later sections, we resolve this tension and argue that behavior of the form \eqref{eq:fidelityMQo} is, in fact, the generic expectation for the basic parent Hamiltonian. We then introduce a modified version of the parent Hamiltonian, Eq. \eqref{eq:modified_Hamiltonian_TFD}, that incorporates the penalty term, Eq. \eqref{eq:penalty_term}. By tuning the penalty coefficient $\xi$ large, the loss of fidelity in the thermodynamic limit can be removed and high-fidelity TFD states can be adiabatically prepared at any system size.

\section{Models}
\label{sec:models}

In this section, we study the ground state of the parent Hamiltonian \eqref{eq:Hamiltonianprop} in a representative set of models:
\begin{itemize}
    \item[(a)] the 1D mixed field Ising model,
    \item[(b)] the 2D transverse-field Ising model,
    \item[(c)] a spin SYK model,
    \item[(d)] a random matrix model.
\end{itemize}
Taken together, these examples probe the proposal across a broad range of physical settings. They include geometrically local models in one dimension (a), which will be our most extensive case study, as well as in two dimensions (b), and also a few-body but geometrically non-local model (c). They moreover provide access to both integrable and non-integrable regimes. The random matrix model (d), in particular, is partly solvable and helps clarify the mechanism of the parent Hamiltonian.

\subsection{Two coupled mixed field Ising chains}
\label{sec:ising}

We study this proposal considering two copies of a 1-dimensional spin chain with $N$ qubits each, governed by the mixed field Ising Hamiltonian with open boundary conditions
\begin{eqnarray}
    H_{{\rm MFI},L/R} &=& J_{z} \sum_{i=1}^{N-1} Z_{i,L/R} Z_{i+1,L/R} + h_x \sum_{i=1}^{N} X_{i,L/R} \nonumber \\
    &&\quad +\, h_z \sum_{i=1}^{N} Z_{i,L/R}\,,
\end{eqnarray}
where $X_i$ and $Z_i$ are single site Pauli matrices.

\begin{figure}[h]
    \centering
    \includegraphics[width=0.48\textwidth]{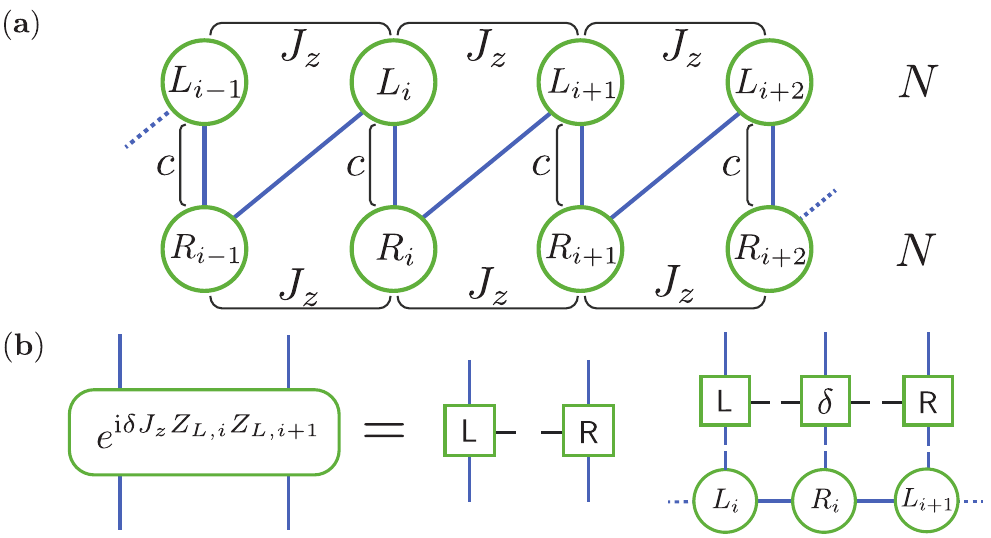}
   \caption{(a) Zigzag ordering used to embed two one-dimensional chains of length $N$ into a single one-dimensional MPS, indicated by the blue lines. The figure shows both the nearest-neighbor Ising terms with coupling $J_z$ and the inter-chain $LR$ couplings with strength $c$ in the coupled Hamiltonian. (b) In the merged-chain representation, applying a Trotterized Ising gate between sites $L_i$ and $L_{i+1}$ requires factorizing the gate into two MPO tensors and inserting an identity MPO, denoted by $\delta$, on the intermediate site $R_i$ induced by the zigzag geometry.}
    \label{fig:zigzagmps}
\end{figure}

In this setup, we choose the $LR$ interaction term in \eqref{eq:Hamiltonianprop} to be ultra-local across the two chains, in the sense that the spin at the $i$-th site of the left chain couples only to the spin at the $i$-th site of the right chain. We take two coupling operators per site, so $K=2N$. The coupled Hamiltonian takes the form:
\begin{eqnarray}\label{eq:coupledHamiltonianMFI}
     &&H  = H_{{\rm MFI},L} + H_{{\rm MFI},R}^*\nonumber \\
    &+& c \sum_{i=1}^N (X_{i,L} - X_{i,R})^2 + c \sum_{i=1}^N (Z_{i,L} - Z_{i,R})^2 \,,
\end{eqnarray}
where the $LR$ coupling $c$ is defined via \eqref{eq:Hamiltonianprop} as $c=J/4$ (we denote it $c$ to distinguish it from the Ising coupling $J_z$). We represent the system in Fig. \ref{fig:zigzagmps}.

Given that the Hamiltonian is short-ranged, we use an MPS ansatz to find the ground state $\ket{\rm GS}$. Concretely, we place the left subsystem on the odd sites of the chain and the right subsystem on the even sites, so that the combined system is represented as a single spin chain with a zigzag ordering, as illustrated in Fig.~\ref{fig:zigzagmps}.  We then determine the ground state by applying the DMRG algorithm to the coupled Hamiltonian \eqref{eq:coupledHamiltonianMFI}. In our simulations, we use a maximum bond dimension \(\chi=625\) and a singular-value cutoff \(\varepsilon=10^{-10}\). We find that the resulting MPS ground states typically have bond dimensions only of order \(\mathcal{O}(10)\), which is well below the imposed maximum. This indicates that the DMRG calculation provides an accurate approximation of the true ground state.

\begin{figure}[h]
\centering
\includegraphics[width=0.48\textwidth]{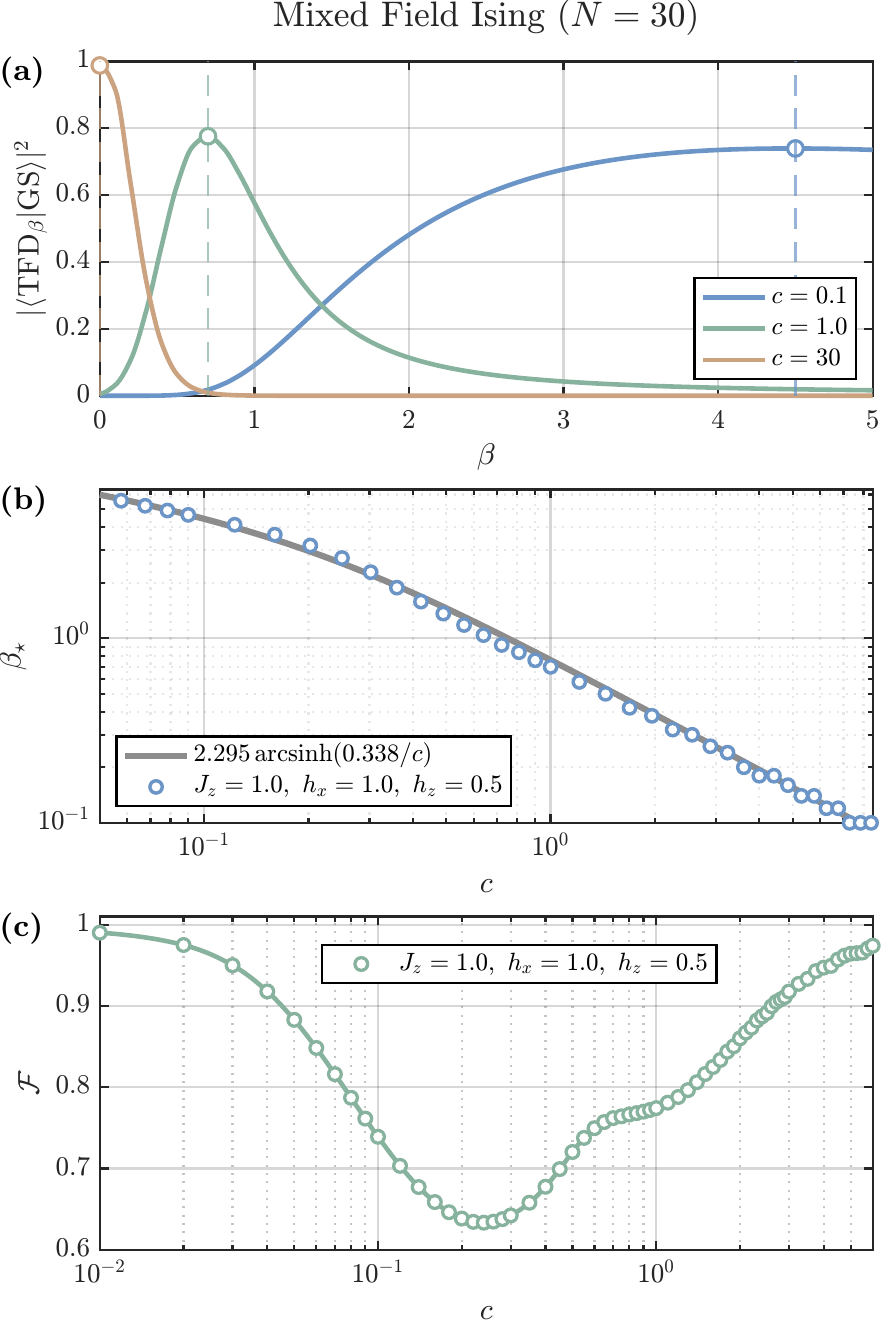}
\caption{Numerical results for two coupled mixed field Ising chains in the chaotic regime, for $N=30$ spins per chain ($60$ spins in total) and parameters $(J_z,h_x,h_z) = (1.0,1.0,0.5)$. (a) Overlap probability $|\langle \mathrm{TFD}_{\beta} | \mathrm{GS}\rangle|^2$ as a function of $\beta$ for three $LR$ couplings, $c=0.1,\,1.0,\,30.0$, with maxima at $\beta_\star \approx 4.3,\,0.7,\,0.1$. (b) Optimal inverse temperature $\beta_\star$ as a function of $c$, showing an approximate inverse relation. (c) Maximum fidelity with the TFD as a function of $c$.}
\label{fig:effective}
\end{figure}

To investigate the overlap between the MPS ground state $\ket{\rm GS}$ and the TFD state, we prepare the latter using imaginary time-evolving block decimation (iTEBD). The MPS TFD state is prepared via second-order Trotterization,
\begin{eqnarray}
    | \mathrm{TFD}_\beta \rangle &=& \left[ \prod_{i=1}^{N} e^{- \frac{\delta}{2} (h_x X_{i,L} + h_z Z_{i,L})} 
    \times \prod_{j=1}^{N-1} e^{- \delta J_z Z_{j,L} Z_{j+1,L}}\right. \nonumber \\
    &\times& \left. \prod_{k=1}^{N} e^{- \frac{\delta}{2} (h_x X_{k,L} + h_z Z_{k,L})} \right]^n | \mathrm{Bell} \rangle ,
\end{eqnarray}
where
\be
    | \mathrm{Bell} \rangle = \bigotimes_{i=1}^N \frac{1}{\sqrt{2}} \left( | \uparrow_{i,L} \uparrow_{i,R} \rangle + | \downarrow_{i,L} \downarrow_{i,R} \rangle \right)\,,
    \label{eq:Bell}
\ee
is the infinite temperature TFD state, and $n=\beta/\delta$ is the number of Trotterization steps. The state $|\mathrm{Bell}\rangle = |\infty\rangle$ is a product of Bell pairs between the spins of the left and right chains. Since all nontrivial Pauli operators act only on the left subsystem, the imaginary-time evolution is applied to one side only; acting on the right instead would yield an equivalent construction. This introduces a Trotterization error of order $\mathcal{O}(n\delta^3)$ in our MPS numerics, and we set $\delta = 0.01$ in most of our numerics. We note that in MPS simulations, we do not expect the Trotterization error of this order of magnitude to dominate other sources of errors, for instance, bond dimension.

Having prepared the ground state of \eqref{eq:coupledHamiltonianMFI} and the TFD, we now present numerical results for their overlap in Fig.~\ref{fig:effective}. In Fig.~\ref{fig:effective}(a), we plot the overlap probability $|\langle {\rm TFD}_\beta|{\rm GS}\rangle|^2$ as a function of the inverse temperature $\beta$. For each value of the $LR$ coupling $c$, we define $\beta_\star(c)$ to be the value of $\beta$ that maximizes the overlap. In Fig.~\ref{fig:effective}(b), we show $\beta_\star(c)$ as a function of $c$ at $N = 30$. We find that $\beta_\star$ quickly saturates to a constant value before we reach $N = 30$ for a given inter-subsystem coupling $c$, and then remains unchanged over the range of system sizes accessible to our numerics beyond $N = 30$. Finally, Fig.~\ref{fig:effective}(c) shows the corresponding fidelity $\mathcal{F} = |\langle {\rm TFD}_{\beta_\star}|{\rm GS}\rangle|^2$ as a function of $c$ for $N=30$ spins per chain. Across the range shown, the overlap remains substantial.

From the general high-temperature analysis for lattice systems in Appendix~\ref{sec:theory}, in particular Eq.~\eqref{eq:betalattice}, we expect the scaling
\be
\beta_\star(c)\sim 1/c\,,
\ee
in good agreement with the numerics.


To understand how this behavior extends to weak $LR$ coupling (low temperatures) in systems with an energy gap, a single spin provides a useful toy model. This is because we expect a dilute gas of excitations at low temperature, which can be crudely modeled as a collection of independent spins. Consider the case in which $H_0 = (1-Z)/2$. With the $-c(X_L X_R+ Z_L Z_R)$ coupling, the coupled ground state lives in the $|00\rangle,|11\rangle$ subspace. The effective Hamiltonian in this subspace is
\begin{equation}
    \begin{bmatrix}
    -c & -c \\ -c & 2-c
\end{bmatrix} = (1 - c) \tilde{I} - \tilde{Z} - c \tilde{X}.
\end{equation}
One can then show that the ground state of this coupled Hamiltonian is exactly a TFD state of $H_0$ with
\begin{equation}
    \beta(c)=2\, \sinh^{-1}\frac{1}{c}.
\end{equation}
Note that we also naturally obtain the expected high energy limit as well, 
\begin{equation}
    \beta(c) \sim \frac{2}{c},\,\ c\to \infty.
\end{equation}

This suggests a fit to the data using a functional form 
\begin{equation}
    \beta(c) \sim A \sinh^{-1} \frac{B}{c}
\end{equation}
where the free parameters $A$ and $B$ offer a slightly more flexible formula that can take into account model-specific features like the energy dispersion of the low-energy excitations. This formula compares well with the data in Fig. \ref{fig:effective}(b). 

\subsection{Exact diagonalization in other models}
\label{sec:other}

We now present exact diagonalization results for the ground state of the parent Hamiltonian in two additional models at small-system size. One of the models is a spin SYK all-to-all model, while the other one is a local $2D$ mixed field Ising model.

In both cases, the ground state of the coupled parent Hamiltonian \eqref{eq:Hamiltonianprop} has a large overlap with an exact TFD state, showing that the mechanism is not specific to the 1D mixed field Ising chain studied in the main text.

\paragraph{Spin SYK---}
We consider a variant of the spin SYK model built from Pauli operators $X,Y$ only \cite{Hanada2024, Swingle2024},
\begin{equation}
H_{\rm Spin\text{-}SYK}
=
\sum_{i_1 < \cdots < i_q}
\;
\sum_{\substack{\alpha_1,\ldots,\alpha_q \in \{x,y\} \\
\#(\alpha=y)\ {\rm even}}}
J^{\alpha_1\cdots\alpha_q}_{i_1\cdots i_q}\,
\sigma^{\alpha_1}_{i_1}\sigma^{\alpha_2}_{i_2}\cdots \sigma^{\alpha_q}_{i_q},
\end{equation}
where $\sigma^{x}_i = X_i$ and $\sigma^{y}_i = Y_i$ are single site Pauli operators. Restricting to terms with an even number of $\sigma^y$ factors ensures that $H_{\rm Spin\text{-}SYK}$ is real in the computational basis, so we may choose the antiunitary map to be complex conjugation, $\Theta=K$. We take the two replicas to be identical for each disorder realization, $H_{0,L}=H_{0,R}=H_{\rm Spin\text{-}SYK}$. Since $KXK=X$ and $KYK=-Y$, the inter-replica coupling takes the form
\begin{equation}
\label{eq:coupling-spin-syk}
H_{LR}
=
c\sum_{i=1}^{N}
\Big[
\left(X_{i,L}-X_{i,R}\right)^2
+
\left(Y_{i,L}+Y_{i,R}\right)^2
\Big].
\end{equation}

\begin{figure}[h]
    \centering
    \includegraphics[width=.48\textwidth]{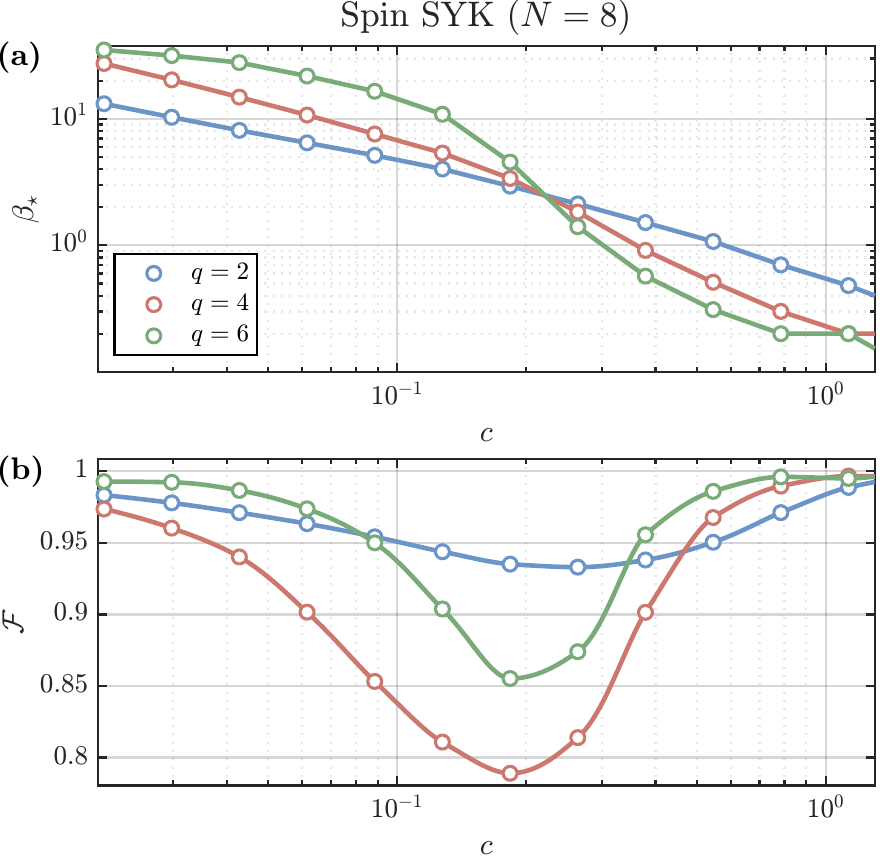}
    \caption{Exact diagonalization results for two coupled spin SYK models with $N=8$ spins per copy ($16$ spins total), averaged over $10$ disorder realizations, as a function of the coupling strength $c$. Panel (a) shows the optimal inverse temperature $\beta_\star$, and panel (b) the corresponding maximal TFD fidelity $\mathcal{F}$.}
    \label{fig:spinsyk}
\end{figure}

The corresponding exact diagonalization results are shown in Fig.~\ref{fig:spinsyk}. Panel (a) shows the optimal inverse temperature $\beta_\star$ as a function of the coupling $c$, while panel (b) shows the corresponding maximal fidelity $\mathcal{F}$ with the TFD state. For $N=8$ spins per copy, averaged over $10$ disorder realizations, the fidelity remains high over a broad range of couplings. This provides a nonlocal, disordered example in which the coupled Hamiltonian selects a high-fidelity TFD ground state.

\begin{figure}[h!]
    \centering
    \includegraphics[width=0.47\textwidth]{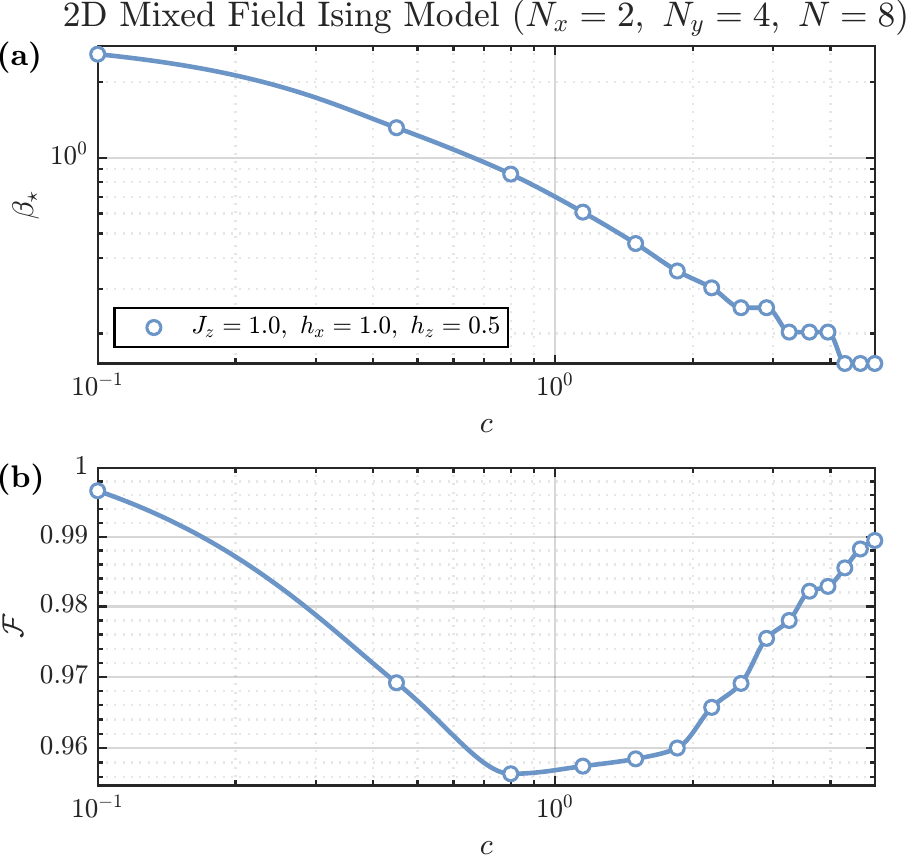}
    \caption{Exact diagonalization results for two coupled 2D mixed field Ising models with $N=8$ spins per copy ($16$ spins total), as a function of the coupling strength $c$. Panel (a) shows the optimal inverse temperature $\beta_\star$, and panel (b) the corresponding maximal TFD fidelity $\mathcal{F}$.}
    \label{fig:2dising}
\end{figure}

\paragraph{2D mixed field Ising model---}
As a higher-dimensional local example, we consider the mixed field Ising model on a $2\times 4$ lattice. Each replica therefore contains $N=N_xN_y=8$ spins, and the two copies are coupled by the same inter-replica interaction used in Eq.~\eqref{eq:coupledHamiltonianMFI}. We again compute, by exact diagonalization, the optimal inverse temperature $\beta_\star$ and the maximal fidelity $\mathcal{F}$ between the ground state of the coupled Hamiltonian and the exact TFD state.

The results are shown in Fig.~\ref{fig:2dising}. As in the spin SYK example, the fidelity remains large over a substantial range of couplings, exceeding $0.95$ in the regime shown. This demonstrates that the parent Hamiltonian construction also works well, at small system sizes, in a geometrically local two-dimensional setting.

\subsection{Random matrix version}
\label{sec:random_matrix_main}

As a complementary example, we also study a random matrix version of the parent Hamiltonian. The point of this model is not to represent a realistic many-body system, but rather to isolate the mechanism of the construction in a setting that is partly analytically solvable. We work on a Hilbert space of dimension $L=2^N$, and take the coupling operators $\{\mathcal O_1,...,\mathcal O_K\}$ to be independent random Hermitian matrices with zero mean and variance normalized so that their matrix elements are of order $L^{-1/2}$. This produces a nonlocal model in which the parent Hamiltonian can be analyzed explicitly for a generic $H_0$.

In the limit of many coupling operators, $K\to\infty$, the interaction $K^{-1} \sum_{\alpha=1}^K (\mathcal{O}_{\alpha,L} -\mathcal{O}^*_{\alpha,R})^2$ self-averages over the random operator ensemble, and the parent Hamiltonian becomes, with unit probability,
\be
H_\infty
=
H_{0,L}+H_{0,R}^* + JN\bigl(\mathbf 1-|\infty\rangle\langle \infty|\bigr),
\ee
where $|\infty\rangle$ is the infinite-temperature TFD. 

This limit is analytically tractable for two reasons. One is that $[H_\infty, H_{0,L} - H_{0,R}^*]=0$, so the parent Hamiltonian preserves the diagonal energy subspace. The off-diagonal subspace is uniformly lifted by an extensive amount $JN$. The second is that, within the diagonal subspace, the parent Hamiltonian can be diagonalized exactly. As we show in Appendix \ref{sec:rmt}, the diagonal eigenstates with energy $E_\lambda = 2\lambda +JN$ are the Choi states of the resolvent of the bare Hamiltonian, $R(\lambda) = (H_0 - \lambda)^{-1}$; explicitly,
\be
|E_\lambda\rangle = |R(\lambda)\rangle \,\propto\, \sum_{n=1}^L\dfrac{1}{\epsilon_n - \lambda}|\epsilon_n\rangle |\epsilon^*_n\rangle\,.
\ee
The value of $\lambda$ is determined by the equation
\be\label{eq:resolventmaincond}
\Tr\frac{1}{H_0-\lambda}=\frac{2L}{JN},
\ee
which has a solution between each pair of neighboring eigenvalues of $H_0$, as well as one additional solution below the lower spectral edge. The ground state is precisely this smallest solution, $\lambda_\star =\min \{\lambda\}$,
\be 
|{\rm GS}\rangle = |E_{\lambda_\star}\rangle\,.
\ee 
More explicitly, if the bare Hamiltonian $H_0$ is bounded below, with smallest energy $\epsilon_{\min}$, then $\lambda_\star \in (-\infty, \epsilon_{\min})$, the ground state is energy $E_{\lambda_\star} <2\epsilon_{\min} + JN$ and it is smaller than the energy of any off-diagonal state.

Thus, in this model, the parent Hamiltonian selects as its ground state a diagonal state in the energy basis, with coefficients controlled directly by the resolvent of $H_0$. A nice consequence is that the overlap with an exact finite-temperature TFD remains $O(1)$ in the large-$N$ limit, as can be checked in different models for $H_0$.

For a GUE choice of $H_0$, the overlap can be obtained analytically in closed form, as we show in Appendix \ref{sec:rmt}. This coarse-grained description of $H_0$ is valid above a critical coupling, $J \geq J_c = 2\sigma/N$, where $\pm 2\sigma$ are the spectral edges, and the states entering \eqref{eq:resolventmaincond} probe the bulk of the spectrum; below this coupling, the discreteness of the spectrum near the edge becomes important. At strong coupling $J/J_c \gg 1$, one moreover finds $\beta_\star \sim 1/J$, matching the same qualitative inverse-coupling behavior seen in the lattice models above.

Moreover, in this model, one can explicitly solve for $\lambda_\star$ and find the gap $\Delta$ of the parent Hamiltonian
\be\label{eq:gaprmtmain}
\Delta = JN\left(1-\frac{J_c^2}{J^2}\right).
\ee
Thus, at fixed $J/J_c$ in the $N\to\infty$ limit, the gap is extensive. This is quite different from what we find below in the mixed-field Ising model, and more generally from what one expects for the parent Hamiltonian in a many-body system; see Appendix \ref{sec:theory} and the infinite-temperature studies \cite{Jian:2022pvj,Guo:2024zmr}. In those cases, the parent Hamiltonian remains thermodynamically gapped, but the gap is typically $O(N^0)$ rather than extensive. The reason is that the coupling operators are very different. In the random matrix model they are fully nonlocal operators acting on the entire Hilbert space, while in physical many-body systems they couple only $O(1)$ degrees of freedom. This extensive gap also explains why the random matrix conclusions are much less sensitive to finite-$K$ effects, as we discuss later in the paper.

\section{Gapped Adiabatic Protocol}
\label{sec:adiabatic}

Having established the structure of the parent Hamiltonian, we now describe an adiabatic protocol \cite{Jansen:2007,Albash2018} for preparing the TFD in the case of the mixed field Ising model. A related adiabatic cooling protocol was proposed in the SYK context in \cite{Bentsen:2023xlu,schuster2025cooling}. 

The starting point is the ground state of the interaction term,
\begin{eqnarray}
    & &H_{LR} = \dfrac{JN}{2K} \sum_{\alpha=1}^K\left(\mathcal{O}_{\alpha,L} -\mathcal{O}_{\alpha,R}^* \right)^2  \nonumber \\
    &=& c \sum_{i=1}^N (X_{i,L} - X_{i,R})^2 + c \sum_{i=1}^N(Z_{i,L} - Z_{i,R})^2 \,.
    \label{eqn: coupling_Hamiltonian}
\end{eqnarray}
Its zero-energy ground state is the infinite temperature TFD, $ |\infty \rangle = |{\rm Bell}\rangle$, described in \eqref{eq:Bell}. We now adiabatically evolve the state from the initial state $| \text{Bell} \rangle$ into the ground state $| \text{GS} (J_z,h_x,h_z,J) \rangle$ of \eqref{eq:Hamiltonianprop} using the time-dependent Hamiltonian
\be
    \mathsf{H}(t)=H_{LR}+s(t)\big(H_{{\rm MFI},L}+H_{{\rm MFI},R}^*\big) \,,
    \label{eqn: adiabatic_Hamiltonian_TFD}
\ee
Here $s(t)$ interpolates slowly from $s(0)=0$ to $s(T)=1$ over a total protocol time $T$.

The adiabatic real time evolution that evolves $| \text{Bell} \rangle$ into $| \text{GS} (J_z,h_x,h_z,J)\rangle$ takes the following form
\be
    U(T) = \mathcal{T} \exp\left(-\mathrm{i} \int_0^T \mathsf{H}(t) \,{\rm d}t \right) \,,
\ee
where $\mathcal{T}$ represents time ordering.

We simulate the resulting real-time adiabatic evolution using a second-order TEBD scheme, with the adiabatic parameter at the $m$-th step chosen to be
\be
    s(m)=\left(m-\frac{1}{2}\right)\frac{\delta t}{T},\qquad m=1,\dots,n\,,
\ee
with $n=T/\delta t$. The corresponding Trotterized unitary for a single time step is
\begin{eqnarray}
    U(s) &=& U_{1\text{-}body}\left(s,h_x,h_z,\frac{\delta t}{2}\right)
    U_{ZZ}\left(s,J_z,\frac{\delta t}{2}\right) \nonumber\\
    &&\times\,
    U_{LR}(c,\delta t)\,
    U_{ZZ}\left(s,J_z,\frac{\delta t}{2}\right)
    \nonumber\\
    &&\times\,
    U_{1\text{-}body}\left(s,h_x,h_z,\frac{\delta t}{2}\right)\,,
    \label{eqn: adiabatic_trotterization_no_penalty}
\end{eqnarray}
and the full adiabatic evolution is obtained by applying a total of $n=T/\delta t$ such layers, updating the value of $s$ at each step, so $U(T) \approx U(\tfrac{(n-1/2)\delta t}{T})...U(\tfrac{3\delta t}{2T}) U(\tfrac{\delta t}{2T})$.

The individual unitaries appearing in Eq.~\eqref{eqn: adiabatic_trotterization_no_penalty} are
\begin{align}
    U_{1\text{-}body} =& \prod_{j=1}^{N} \exp\!\left[-\mathrm{i} s \frac{\delta t}{2} \left(h_x X_{j,L} + h_z Z_{j,L}\right)\right] \nonumber\\
    &\times \prod_{k=1}^{N} \exp\!\left[-\mathrm{i} s \frac{\delta t}{2} \left(h_x X_{k,R} + h_z Z_{k,R}\right)\right]\,, 
\end{align}
\begin{align}
    U_{ZZ} = &\prod_{i=1}^{N-1} \exp\!\left[-\mathrm{i} s \frac{\delta t}{2} J_z Z_{i,L} Z_{i+1,L}\right] \nonumber\\
    &\times \prod_{k=1}^{N-1} \exp\!\left[-i s \frac{\delta t}{2} J_z Z_{k,R} Z_{k+1,R}\right]\,, 
\end{align}
\begin{align}
    U_{LR} =& \prod_{i=1}^{N} \exp\!\left[\mathrm{i}\, 2c\, \delta t\, Z_{i,L} Z_{i,R}\right] \nonumber\\
    &\times \prod_{k=1}^{N} \exp\!\left[\mathrm{i}\, 2c\, \delta t\, X_{k,L} X_{k,R}\right]\,.
\end{align}

The midpoint choice of the adiabatic parameter $s(m)$ ensures that each second-order TEBD layer has a Trotter error of order $\mathcal{O}(\delta t^3)$. We also note that the terms within each of the unitaries $U_{1\text{-}body}$, $U_{ZZ}$, and $U_{LR}$ commute among themselves, which makes this decomposition particularly convenient for implementation. 

\subsection{Performance and timescales}

\begin{figure}[h]
\centering
\includegraphics[width=0.48\textwidth]{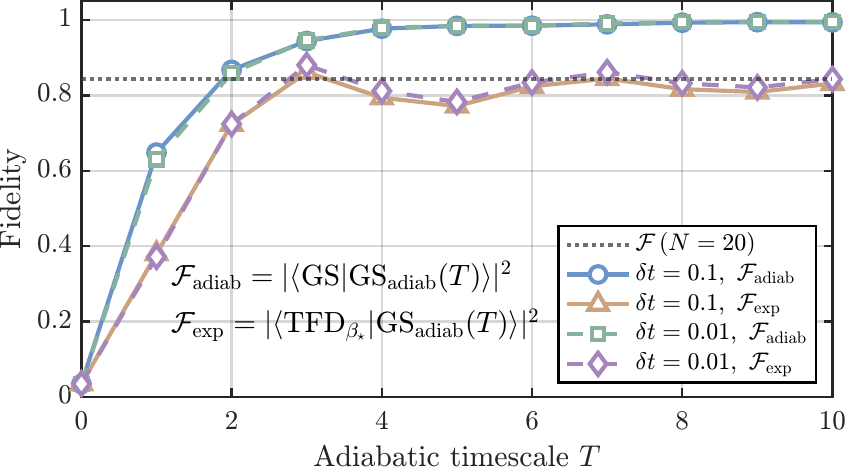}
\caption{Fidelities used to benchmark the performance of the adiabatic protocol as a function of the total evolution time $T$. The parameters are $(J_z,h_x,h_z,c)=(1.0,1.0,0.5,1.0)$. In the MPS simulation, we use a bond dimension $\chi=100$ and a singular value cutoff $\varepsilon=10^{-10}$.}
\label{fig:adiabatic}
\end{figure}

We report the numerical performance of the adiabatic protocol as a function of the total evolution time $T$ for one representative choice of parameters, $(J_z,h_x,h_z,c)=(1.0,1.0,0.5,1.0)$, and $N=20$ spins per chain. We monitor the adiabatic state $\ket{ {\rm GS}_{\rm adiab}(T)} = U(T) \ket{\rm Bell}$ and consider the time evolution of the fidelities, which benchmark how close the adiabatic state is to the true ground state and to the TFD
\begin{gather}
\mathcal{F}_{\rm adiab}(T) = |\bra{\rm GS}\ket{{\rm GS}_{\rm adiab}(T)}|^2  \,,\\[.2cm]
\mathcal{F}_{\rm exp}(T) = |\bra{{\rm TFD}_{\beta_\star}}\ket{{\rm GS}_{\rm adiab}(T)}|^2\,.
\end{gather}
In this regime of parameters, the TFD has an inverse temperature $\beta_\star \approx 0.69$. We consider two different Trotter steps, so that each time step has a circuit depth $10$ or $100$. The numerical results are shown in Fig. \ref{fig:adiabatic}.

The initial point at $T=0$ shows the fidelities of the initial state of the adiabatic protocol, $\ket{{\rm GS}_{\rm adiab}(0)}$. The fidelity $\mathcal{F}_{\rm adiab}(T)$ exceeds $0.9$ around $T=3.0$ and begins to saturate above $0.98$ near $T=4.0$. This implies that, for the chosen Ising parameters, this adiabatic protocol time is sufficient to produce a high-fidelity ground state. 

Another quantity of interest is the fidelity of the adiabatic state with the TFD, $\mathcal{F}_{\rm exp}(T)$. The constant value $\mathcal{F} \approx 0.84$ in Fig.~\ref{fig:adiabatic} represents the fidelity of the exact ground state of the coupled Hamiltonian with the TFD of interest and therefore serves as a benchmark for the best performance one can expect in the optimal case. We see that this fidelity is also reached once the protocol enters the adiabatic regime around $T\approx 3.0$, after which it only slightly fluctuates around the benchmark value.

A final feature worth emphasizing is that the quality of the adiabatic protocol is affected much more by the total protocol time $T$ than by the Trotter step size $\delta t$, provided $\delta t$ is chosen sufficiently small. In particular, decreasing the Trotter step from $\delta t=0.1$ to $\delta t=0.01$ leads only to a minor improvement in the fidelity $\mathcal{F}_{\rm exp}$. This has an important implication for the design of near-term quantum algorithms aimed at preparing high-fidelity TFD states in the mixed field Ising chain. Already at $T=3.0$ with Trotter step $\delta t=0.1$, we obtain
\be
\mathcal{F}_{\rm adiab}(3.0) \approx 0.94,
\qquad
\mathcal{F}_{\rm exp}(3.0) \approx 0.86,
\ee
with the latter already close to the optimal benchmark. This suggests that a high-quality TFD-like state can be prepared using as few as 30 layers of second-order Trotterized unitaries in Eq.~\eqref{eqn: adiabatic_trotterization_no_penalty}, making the protocol promising for near-term quantum implementations.\\

\begin{figure}[h]
\centering
\includegraphics[width=0.48\textwidth]{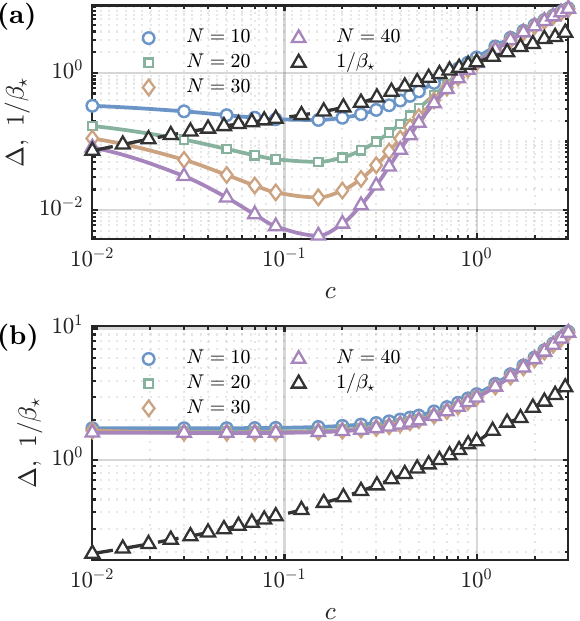}
\caption{Gap $\Delta$ and inverse optimal temperature $1/\beta_\star$ versus the $LR$ coupling $c$ for the coupled mixed field Ising chains. The first excited state is numerically obtained with DMRG, with a large energy penalty imposed on the subspace parallel to the ground state. (a) Chaotic regime $(J_z,h_x,h_z)=(1.0,1.0,0.5)$. (b) Paramagnetic regime $(J_z,h_x,h_z)=(1.0,1.8,0.1)$. The main difference between the two cases is that the gap of $H_0$ is larger in the paramagnetic regime.}
\label{fig:gap}
\end{figure}

\paragraph{Timescales ---} The timescale of the adiabatic protocol is controlled by the smallest gap of the coupled Hamiltonian along the interpolation path, together with the smoothness of the schedule, in accordance with the adiabatic theorem \cite{Jansen:2007,Albash2018}. In Fig.~\ref{fig:gap}, we plot the gap $\Delta$ as a function of the $LR$ coupling $c$ for different values of $N$ and in different regimes of the mixed field Ising chains. Broadly speaking, at large coupling $c$, the gap enters a temperature-controlled regime and therefore remains non-extensive, while at small coupling it approaches the gap of the Ising Hamiltonian $H_{\rm MFI}$. It follows that if the original Hamiltonian $H_{\rm MFI}$ is thermodynamically gapped, the adiabatic timescale is also expected to remain non-extensive, $O(1/\Delta)$, for any value of the TFD temperature we want to prepare in the lab.

\section{Scalability problem: fidelity loss}
\label{sec:scaling-problem}

\begin{figure}[h]
\centering
\includegraphics[width=0.48\textwidth]{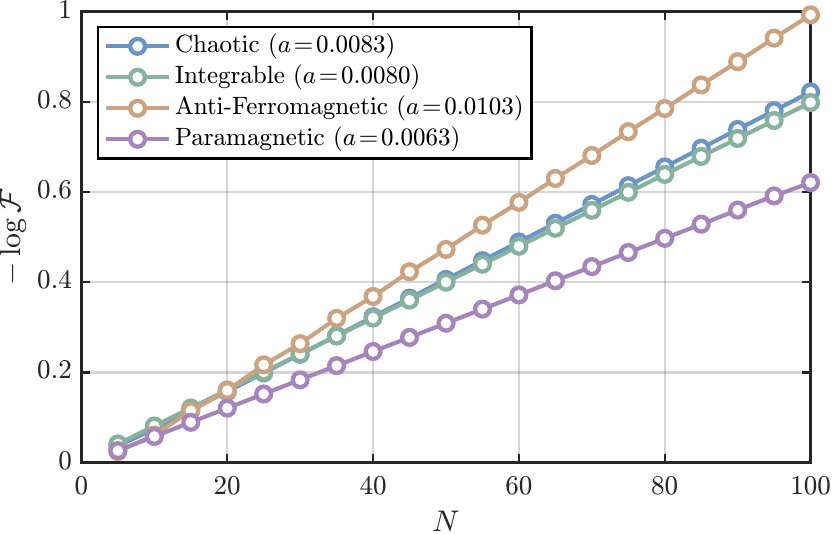}
\caption{Fidelity decay with system size in four representative regimes of the mixed field Ising model. The parameter values are $(J_z,h_x,h_z)=(1.0,1.0,0.5)$ for the chaotic regime, $(1.0,1.0,0.0)$ for the integrable regime, $(1.0,0.3,0.1)$ for the antiferromagnetic regime, and $(1.0,1.8,0.1)$ for the paramagnetic regime. The approximately linear behavior $-\log \mathcal{F} \approx aN$ indicates an exponential decay of the fidelity with system size, even though the slope $a$ is small. }
\label{fig:neglogF}
\end{figure}

We now test the proposal numerically at larger system sizes in the coupled mixed field Ising chains, reaching up to $N=100$ spins per chain ($200$ in total). As shown in Fig.~\ref{fig:neglogF}, the ground-state fidelity with the TFD is well fit by an exponential decay with system size,
\be\label{eq:expdecayfid}
\mathcal F \approx e^{-aN},
\ee
where the coefficient $a$ depends on both the Ising parameters $(J_z,h_x,h_z)$ and the inter-chain coupling $c$. This behavior clearly departs from the general proposal of Cottrell et al. \cite{Cottrell_2019} in Eq.~\eqref{eq:fidelityprop}. At the same time, the fitted exponent is numerically small, typically in the range $\tfrac{1}{160}\lesssim a \lesssim \tfrac{1}{100}$, so the fidelity remains very high throughout the range $N\lesssim 50$ accessible to many practical applications.

In Appendix~\ref{sec:theory}, we trace the failure of the general proposal of Cottrell et al.~\cite{Cottrell_2019} in Eq.~\eqref{eq:fidelityprop} to the assumption that the ground state $\ket{\rm GS}$ is fully supported on the diagonal energy subspace, namely the subspace annihilated by $H_{0,L}-H^*_{0,R}$. This is true only in the limit of sufficiently many coupling operators, assuming the operators behave independently in the ETH sense. At finite $K$, fluctuations in the interaction generate admixture into the off-diagonal subspace. Since the TFD lies entirely in the diagonal sector, any such leakage necessarily lowers the fidelity.

A simple perturbative estimate gives
\be
\|P_{\rm off}\ket{{\rm GS}}\|^2
\sim
\frac{(JN)^2}{K\Delta_{\rm off}^2},
\ee
where
\be
P_{\rm off}
=
\sum_{n\neq m}
\ket{\epsilon_n}\ket{\epsilon_m^*}\bra{\epsilon_n}\bra{\epsilon_m^*}
\ee
is the projector onto the off-diagonal energy subspace, and $\Delta_{\rm off}$ is the gap between the $K=\infty$ diagonal ground state and the off-diagonal states. The diagonal analysis is therefore perturbatively controlled only if
\be
K\gg \left(\frac{JN}{\Delta_{\rm off}}\right)^2.
\ee
For local spin systems, it is natural to expect $\Delta_{\rm off}$ to be of the same order as the ordinary gap of the parent Hamiltonian, which is non-extensive. Under this assumption, suppressing the leakage would require at least $K\gtrsim O(N^2)$. In the mixed field Ising chain studied here, however, we only have $K=2N$, so this condition is not satisfied. It is also possible that the ETH-independence assumption itself becomes inaccurate, but in either case, the conclusion is the same: $1/K$ corrections to the fidelity are not perturbatively small.

This gives a natural explanation for the behavior in Fig.~\ref{fig:neglogF}. The ground state is no longer confined to the diagonal energy sector but spreads into a much larger off-diagonal space. Since the diagonal sector occupies only an exponentially small fraction of the full double-copy Hilbert space, one should then expect the overlap with the TFD to decrease exponentially with system size, in agreement with the numerics.

\paragraph{Random matrix version---}
The random matrix version of the parent Hamiltonian provides a useful contrast because finite-$K$ effects are much milder in that fully nonlocal setting. There one again finds perturbatively
\be
\|P_{\rm off}|{\rm GS}\rangle\|^2 \sim \frac{(JN)^2}{K\Delta_{\rm off}^2},
\ee
but now the relevant gap $\Delta_{\rm off}/J = O(N)$ is itself extensive, just like the diagonal gap (see, for example, Eq.~\eqref{eq:gaprmtmain}). As a result, the off-diagonal leakage is already suppressed for $K=O(N^0)$, so a small number of random coupling operators is enough to keep the ground state close to the $K=\infty$ resolvent state and preserve an $O(N^0)$ overlap with the optimal TFD.  We present finite-$K$ numerics of the fidelity loss in Appendix~\ref{sec:rmt}, where this behavior is confirmed.

\section{Improved Parent Hamiltonian with penalty}
\label{sec:penalty-term}

With the motivation of making the ground state diagonal, we introduce the simplest off-diagonal ``penalty term'' in the Hamiltonian \eqref{eq:Hamiltonianprop} and instead consider the improved parent Hamiltonian
\begin{eqnarray}
    H &=& H_{0,L} + H^*_{0,R} + \dfrac{JN}{2K} \sum_{\alpha=1}^K\left(\mathcal{O}_{\alpha,L} -\mathcal{O}_{\alpha,R}^* \right)^2 \nonumber \\
    &+& \frac{\xi}{2N} (H_{0,L} - H_{0,R}^*)^2 , 
    \label{eq:modified_Hamiltonian_TFD}
\end{eqnarray}
for some coupling $\xi$ with units of inverse energy. We choose the normalization of the penalty term so that the trace of the penalty term is extensive in $N$. This normalization is only a matter of convention and may be traded for a corresponding redefinition of $\xi$.

\begin{figure}[h]
\centering
\includegraphics[width=0.45\textwidth]{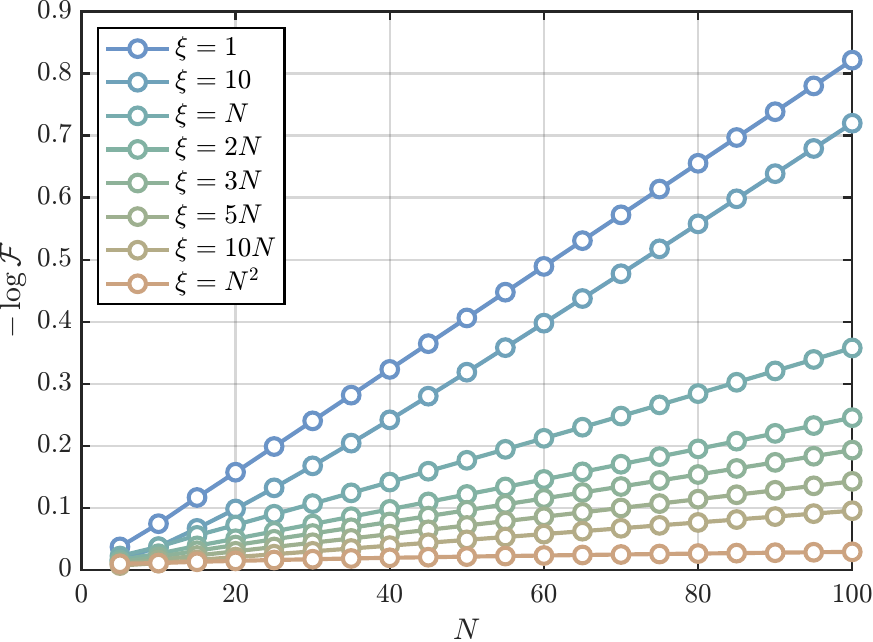}
  \caption{Ground state fidelity with the TFD as a function of system size $N$ for two coupled mixed field Ising chains. The data is taken near the maximally chaotic regime, with parameters $(J_z,h_x,h_z,c)=(1.0,1.0,0.5,1.0)$. The near-linear dependence of $-\log \mathcal{F}$ on $N$ signals an exponentially decaying overlap. Increasing the penalty strength $\xi$ reduces the slope of this decay.}
\label{fig:fidwithpenalty}
\end{figure}

\begin{figure}[h]
\centering
\includegraphics[width=0.45\textwidth]{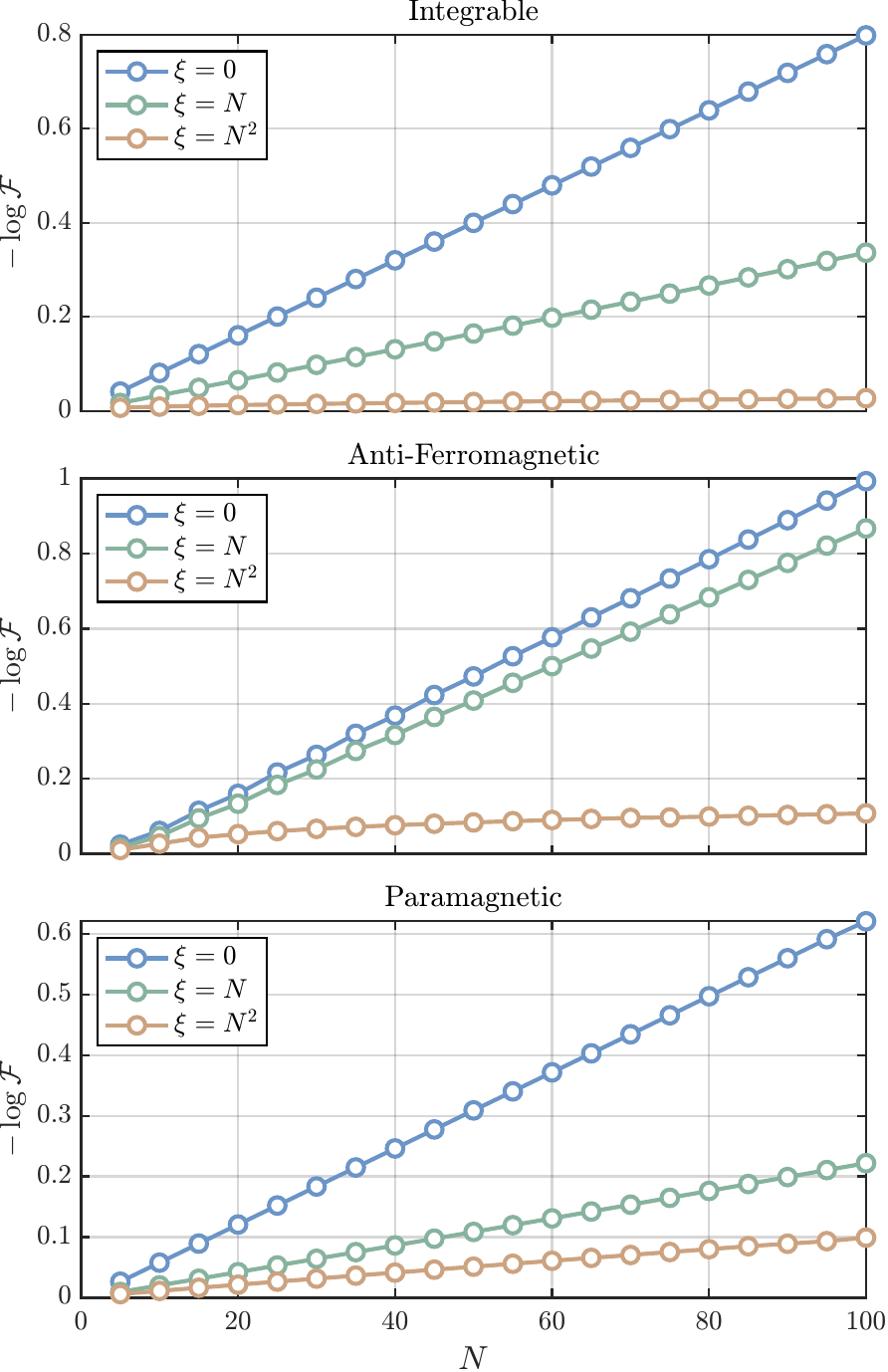}
\caption{Ground state fidelity with the TFD $-\log \mathcal{F}$ as a function of subsystem size $N$ for two coupled mixed field Ising chains with penalty term. The three panels correspond to the integrable regime ($J_z=h_x=1.0$, $h_z=0.0$), the antiferromagnetic regime ($J_z=1.0$, $h_x=0.3$, $h_z=0.1$), and the paramagnetic regime ($J_z=1.0$, $h_x=1.8$, $h_z=0.1$).}
\label{fig:regimes}
\end{figure}

Numerically, we study the effect of the penalty term in \eqref{eq:modified_Hamiltonian_TFD} for the coupled mixed field Ising chains. Fig.~\ref{fig:fidwithpenalty} shows the detailed behavior in the near-maximally chaotic regime, while Fig.~\ref{fig:regimes} shows the corresponding behavior for representative values of the penalty strength $\xi$ in three additional regimes: integrable, antiferromagnetic, and paramagnetic. In these calculations, we obtain the ground state of the modified Hamiltonian using an MPS ansatz and the DMRG algorithm, with a maximum bond dimension $\chi=625$ and a singular-value cutoff $\epsilon=10^{-10}$. The TFD state is again prepared using imaginary-time TEBD.

In all cases, $-\log \mathcal{F}$ grows approximately linearly with $N$, indicating that the overlap decays exponentially with system size. Increasing the penalty strength $\xi$ (measured in units of $1/J_z$) reduces the corresponding decay exponent, which appears as a smaller slope in these plots. In the near-maximally chaotic regime, the $\xi = N^2$ curve has an extremely small slope over the accessible system sizes shown, and larger-system numerics would be needed to determine whether the fidelity saturates as the system size increases. We leave this question for future work, although we turn to an estimate that indicates that the fidelity loss would certainly be arrested for $\xi \gtrsim N^3$.

\subsection{Suppression of off-diagonal leakage}

At large enough $\xi$, the penalty term has a simple perturbative effect. As explained in Appendix~\ref{sec:theory}, we write
\be
H = H_{\infty} + \frac{\xi}{2N}(H_{0,L}-H_{0,R}^*)^2 + \delta H ,
\ee
where $H_{\infty}$ is the $K\to\infty$ operator-averaged Hamiltonian, assuming the coupling operators behave as independent draws in the ETH sense, and $\delta H$ is the finite-$K$ fluctuation. Its characteristic size is $\delta H \sim \frac{JN}{\sqrt K}$.

The key point is that both $H_{\infty}$ and the penalty term commute with
$H_{0,L}-H_{0,R}^*$, so they preserve the diagonal subspace. The $K=\infty$
ground state is expected to lie in this diagonal subspace, as discussed in
Appendix~\ref{sec:theory}.

The penalty term vanishes on the diagonal subspace, but it raises the energy of an off-diagonal state $\ket{\epsilon_n}_L\ket{\epsilon_m^*}_R$ with $n\neq m$ by
\be
\frac{\xi}{2N}(\epsilon_n-\epsilon_m)^2.
\ee
The finite-$K$ fluctuation $\delta H$ can still mix the diagonal and off-diagonal sectors, but this mixing is suppressed because the penalty term raises the energy of off-diagonal states. Parametrizing the off-diagonal sector by the relative energy $\omega = \epsilon_n-\epsilon_m$, standard perturbation theory gives the estimate
\be
\|P_{\rm off}|{\rm GS}\rangle\|^2
\;\sim\;
\frac{N^2}{K}
\int d\omega\,
\frac{\hat{\rho}_{\rm off}(\omega)}
{\big(\Delta_{\rm off}+\xi \omega^2/N\big)^2},
\label{eq:offdiagweightpenalty}
\ee
where $\rho_{\rm off}(\omega)\equiv \sum_{a\in{\rm off}} |\langle a|\delta H|0\rangle|^2 \delta(\omega-\omega_a)$ is the off-diagonal spectral function generated by $\delta H$, and $\Delta_{\rm off}$ is the $K=\infty$ gap to the off-diagonal sector at $\xi=0$. Here $\hat \rho_{\rm off}(\omega) = K\rho_{\rm off}(\omega)/N^2$ is the $O(1)$ spectral function.

For few-body coupling operators, we expect $\Delta_{\rm off}$ to be non-extensive in $N$, much like the diagonal gap itself. The required scaling of $\xi$ then depends on the low-frequency behavior of $\rho_{\rm off}(\omega)$. If $\rho_{\rm off}(\omega)$ approaches a constant as $\omega\to 0$, the integral is dominated by
\be
\omega \sim \sqrt{\frac{N\Delta_{\rm off}}{\xi}},
\ee
which gives
\be
\|P_{\rm off}|{\rm GS}\rangle\|^2
\;\sim\;
\frac{N^{5/2}}{K\,\sqrt{\xi}\,\Delta_{\rm off}^{3/2}}
\ee
up to an overall prefactor. Therefore, if $K\sim N$, this estimate suggests the scaling
\be
\xi \gtrsim N^3,
\ee
up to model-dependent factors and powers of $\Delta_{\rm off}$. The precise power of $N$ depends on the low-frequency behavior of $\rho_{\rm off}(\omega)$: stronger low-frequency weight pushes the required scaling upward, while a faster decay at small $\omega$ makes the penalty term more effective.

\subsection{Improving the TFD fidelity of the Maldacena-Qi wormhole}
\label{sec:syk_main}

The effects of the penalty term can be computed explicitly for the case of the holographic wormhole ground state \cite{maldaqi}, obtained by coupling two SYK systems through the Hamiltonian \eqref{eq:Hamiltonianprop}. We summarize the main point here and leave the details to Appendix~\ref{sec:SYK}.

At leading order in the $LR$ coupling, the Maldacena-Qi interaction simply changes the effective action for the reparametrization modes. At this level, the wormhole ground state is still described by a TFD, so the fidelity is unity. The first deviation arises at subleading order in the interaction, from quench diagrams associated with bulk particle creation. These processes take the state slightly out of the diagonal sector and therefore reduce its overlap with the TFD. The result is an exponentially small-in-$N$ correction to the fidelity,
\be\label{eq:fidelmainMQ}
\log \mathcal F \sim -N \times (\text{small coefficient}),
\ee
with the coefficient parametrically suppressed at large $q$, for $q$-body SYK.

Adding the penalty term suppresses such contributions by lifting configurations with nonzero relative boost energy between the two boundaries. As a result, the quench diagrams responsible for fidelity loss are suppressed, especially at large Euclidean separations. The fidelity loss coefficient is correspondingly reduced
\be\label{eq:fidelmainMQpen}
\left.\log \mathcal F \;\right|_{\xi}\sim -N \times (\text{small coefficient})\times (N/\xi)^{2-4\Delta_{\mathcal O}},
\ee
where $\Delta_{\mathcal O}<1/2$ is the infrared conformal dimension of the interaction operator. For $4$-body SYK and single Majorana fermion coupling, $\Delta_{\mathcal O} =1/4$.

The key conclusion is that, in the Maldacena-Qi wormhole, an appropriate $N$-dependent penalty strength is sufficient to restore $O(N^0)$ fidelity at large $N$. The scaling implied by \eqref{eq:fidelmainMQpen} is
\be\label{eq:mqcouplingmain}
\mathcal{J}\xi \gtrsim N^{\frac{3-4\Delta_{\mathcal O}}{2-4\Delta_{\mathcal O}}}\,.
\ee
For $\Delta_{\mathcal O}=1/4$, this gives the quadratic scaling $\mathcal J \xi \gtrsim N^2$. For large $q$, where $\Delta_{\mathcal O}=1/q$, \eqref{eq:mqcouplingmain} gives $\mathcal J \xi \gtrsim N^{3/2}$. More generally, this is consistent with the spectral-function estimate in \eqref{eq:offdiagweightpenalty}: in the Maldacena-Qi wormhole, the low-frequency off-diagonal weight is not constant, but instead depends on $\Delta_{\mathcal O}$, which changes the power of $N$ required for the penalty term to be effective. As the interaction becomes marginal, $\Delta_{\mathcal O}\to 1/2$, the exponent diverges, indicating that particle production cannot be suppressed by any polynomial scaling of the penalty strength.

\section{Outlook}
\label{sec:outlook}

We have shown that simple coupled parent Hamiltonians can produce TFD-like ground states and have used this to propose concrete adiabatic protocols for TFD preparation. We also showed explicitly that these adiabatic protocols perform well in practice and should be accessible on current or near-term quantum devices. While the fidelity with the exact TFD remains high at system sizes accessible to current quantum devices, it generically decays exponentially at larger system sizes due to leakage into the off-diagonal energy sector. We addressed this scalability problem by introducing an energy-mismatch penalty term and showed analytically and numerically that it can substantially improve the ground state overlap.

Even without the penalty term, the ground state of the parent Hamiltonian may still capture the physics of interest. For example, if we are interested in some universal property of a phase of matter, then as long as the ground state is in the same phase as the true TFD, the universal physics is accessible. If we are interested in some intensive but non-universal physical property, such as the energy density or magnetic susceptibility, then we also expect that the ground state will yield a result that has bounded error (not growing with system size) compared to the ground truth obtained from the TFD state. It will be interesting and important to investigate these expectations in more detail across a variety of models.

If a high overlap with the true TFD state is crucial, then we showed that one may include the penalty term with a large coefficient to achieve it. Since the gap of the parent Hamiltonian is only expected to increase with the addition of the penalty term, an adiabatic state preparation algorithm is still available. Moreover, since we argued that $\xi$ needs to scale at worst as a polynomial in $N$, it follows that the adiabatic algorithm can prepare a high overlap state in poly($N$) depth, although the algorithm with penalty term may not be accessible on near-term devices. It is also interesting to explore other ways to effectively project onto the diagonal energy subspace, such as applying phase estimation to $H_{0,L} - H_{0,R}$.

The domain of applicability of our method also deserves further study. We conjecture that the three conditions outlined in the introduction are sufficient for the method to function, but it is not clear whether they are necessary or if they can be further simplified. For example, when the mixed field Ising chain is tuned to integrability, where ETH presumably fails, the ground state of the parent Hamiltonian still approximately tracks the true TFD. It is therefore interesting to study a wider variety of models that fail one or more of our criteria to explore the full scope of the method. Moreover, it is well known that simple coupled Hamiltonians of this type suffice to prepare exact TFD states for non-interacting fermion~\cite{Swingle_2016_mixed_s} and boson systems. Additionally, it is particularly interesting to explore what happens to the parent Hamiltonian when the target TFD state approaches a thermal phase transition.

Another aspect of our procedure is the choice of the $\mathcal{O}$ coupling operators. We gave a simple choice that worked across the examples studied here, but it is interesting to study other choices including how the overlap with the TFD depends on $K$. For example, we initially tried to include just $\mathcal{O} = Z$ couplings in the Ising model, but we found this led to poor overlaps. One principle is thus that the infinite temperature thermofield double state should be the unique ground state of the $J \to \infty$ Hamiltonian. It is also important to take symmetry into account. For example, if $H_0$ has a unitary symmetry $G$, then the TFD typically spontaneously breaks the $G \times G$ symmetry of $H_{0,L} + H_{0,R}^*$ down to its diagonal subgroup. Then, in order for the TFD to be the unique ground state of the parent Hamiltonian, it must be that the parent Hamiltonian explicitly breaks the $G \times G$ symmetry via the coupling terms (otherwise the ground state would need to be degenerate in the thermodynamic limit as it spontaneously breaks a symmetry of the parent Hamiltonian). Other constraints arise when thinking about the low temperature limit. If the target Hamiltonian has scale invariant behavior at low energy, then we expect that the $\mathcal{O}$s need to be chosen to be relevant operators in the renormalization group sense. To reach thermal states above a gapped ground state, the coupling operators may need to have a non-vanishing expectation value at low energy. The principle in both cases is to make sure that the effect of the coupling doesn't vanish too rapidly at low energy.

At this point it is useful to mention a related problem wherein we view the TFD state as a variational ansatz for the parent Hamiltonian ground state. This is a different optimization than the one we considered in the bulk of this paper, but when the max overlap with the TFD state is high, these two problems give approximately the same answer. A desirable feature of the variational problem is that the variational energy is very simple, involving just the thermal energy of the one-sided Hamiltonian and a particular thermal 2-point function of the coupling operators $\mathcal{O}$. We want to further investigate the variational problem and its relationship to the main problem we studied. For example, under what conditions does the variational inverse temperature $\beta_{\text{var}}$ track $\beta_\star$? As one application, by choosing the coupling operators to be irrelevant, say $(p+1)$-fermion $\mathcal{O}$s in the $p$-body SYK model, we conjecture that it is possible to construct a $\beta_{\text{var}}(J)$ which jumps as a function of $J$. It is interesting to study if the same phenomenon occurs in $\beta_\star(J)$.

Finally, we emphasize that it is not necessary to know the thermodynamics of the model of interest in advance. One approach is to use an analog of thermometry, say by coupling the target system to another small probe system with a well calibrated relationship between energy density and temperature. By letting them equilibrate, one can read off the temperature of the target system from the probe. Another approach is to approximately map out the thermodynamics as follows. Suppose we have a quantum device that prepares the ground state of the parent Hamiltonian \eqref{eq:Hamiltonianprop} for a given $J$. From this state, we may measure both the average one-sided energy $\langle H_{0,L}\rangle$ and its variance $\Delta H_{0,L}^2 = \langle (H_{0,L} - \langle H_{0,L}\rangle)^2\rangle$. Viewing $\langle H_{0,L}\rangle$ as a function of $\beta = \beta_\star(J)$, the chain rule gives $\frac{d \langle H_{0,L}\rangle}{d J} = \frac{d \langle H_{0,L} \rangle}{d \beta} \frac{d\beta_\star}{d J}$. If the state is close to a TFD, then we have $\frac{d \langle H_{0,L} \rangle}{d \beta} \approx - \Delta H_{0,L}^2$ and we can map out $\beta_\star(J)$ by integrating $- \frac{1}{\Delta H_{0,L}^2}\frac{d \langle H_{0,L}\rangle}{dJ} = \frac{d \beta_\star(J)}{d J}$ down from $J=\infty$.\\

\textbf{Acknowledgments} - We acknowledge support from the U.S. Department of Energy through GeoFlow DE-SC0019380 (M. Sasieta, B. Swingle) and DE-SC0025934 (B. J. J. Khor), and from the Heising-Simons Foundation through grant 2024-4849 (N. LiTenn, B. Swingle). This work is supported in part by the Leinweber Institute for Theoretical Physics (M. Sasieta). We thank B. Freivogel and B. Kobrin for discussions. B. Swingle also thanks Claude Opus 4.7, GPT 5.5, Gemini 3.1, and prior models for feedback on the draft and the physics; all the writing and calculations were carried out by the authors.

\bibliographystyle{ourbst}
\bibliography{bibliography}

\vspace{.2cm}

\onecolumngrid

\appendix

\vspace{1cm}
{\centering
\large\bfseries
Supplementary Material
\par}

\begin{center}
    {\large\bfseries Contents}
\end{center}
\vspace{0.5em}

\noindent
\hyperref[sec:theory]{Appendix A. General aspects of the ground state in chaotic systems}
\hfill \pageref{sec:theory}

\noindent
\hyperref[sec:rmt]{Appendix B. Ground state in random matrix theory}
\hfill \pageref{sec:rmt}

\noindent
\hyperref[sec:SYK]{Appendix C. Details of the Maldacena-Qi wormhole calculations}
\hfill \pageref{sec:SYK}

\section{General aspects of the ground state in chaotic systems}
\label{sec:theory}

In this appendix, we revisit the proposal of Cottrell et al. \cite{Cottrell_2019} with two main objectives. First, we clarify the structure of the ground state and place its analysis on firmer footing, extending the discussion to physical settings relevant to this paper, including lattice systems. Second, we explain why the conclusion regarding the fidelity with the TFD does not follow from the assumptions of the argument once these are examined more carefully.

The first and rather mild assumption is that the Hamiltonian $H_0$ is quantum chaotic in the sense that it satisfies the eigenstate thermalization hypothesis (ETH) \cite{PhysRevA.43.2046,Srednicki:1994mfb}. In this case, for a ``simple'' few-body operator $\mathcal{O}$, the matrix $\mathcal{O}_{nm}:=\bra{\epsilon_n}\mathcal{O}\ket{\epsilon_m}$ can be modeled as an instance of an ensemble of banded random matrices
\be\label{eq:ETHform} 
\mathcal{O}_{nm} = \sqrt{\dfrac{g(\overline{\epsilon}_{nm},\omega_{nm})}{\varrho(\overline{\epsilon}_{nm})}}\,R_{nm}\,,
\ee 
with mean and variance
\be\label{eq:ETHaverage}
\overline{R_{nm}}=0\,,\qquad \overline{R_{nm}R_{pq}^*}= \delta_{np}\delta_{mq}\,,
\ee
for $\overline{\epsilon}_{nm} = (\epsilon_n + \epsilon_m)/2$, $\omega_{nm} = \epsilon_n -\epsilon_m$. Here $\varrho(\epsilon)$ is the thermodynamic density of states, and $g(\epsilon,\omega)$ is the smooth envelope function that determines the microcanonical real-time two-point function of the operator by Fourier transform in $\omega$. For simplicity, we assume that the microcanonical one-point function of the operators vanishes.

The operators $\{\mathcal{O}_\alpha\}_{\alpha=1}^K$ entering the Hamiltonian \eqref{eq:Hamiltonianprop} can then be regarded as independent instances in the ETH sense, and for simplicity we will moreover assume that all of them possess the same envelope function. The second assumption underlying the analysis of \cite{Cottrell_2019} is that one can, in the $K\to \infty$ limit, replace the interaction term of the Hamiltonian by its ETH average
\be 
\lim_{K\to\infty}\dfrac{1}{K}\sum_{\alpha=1}^K\left(\mathcal{O}_{\alpha,L} -\mathcal{O}_{\alpha,R}^* \right)^2 = \overline{\left(\mathcal{O}_{L} -\mathcal{O}_{R}^*\right)^2}\,.
\ee 
This assumption is, however, not satisfied in the treatment of \cite{Cottrell_2019}, where $K$ is instead taken to be small, in fact non-extensive. For finite $K$, as we will see, there are additional corrections suppressed by $1/\sqrt{K}$, which nevertheless induce important effects and make the analysis of \cite{Cottrell_2019} invalid as far as the fidelity is concerned.\\

\paragraph{Energy shift ---} Expanding $(\mathcal O_L-\mathcal O_R^*)^2=\mathcal O_L^2+\mathcal O_R^{*2}-2\,\mathcal O_L\mathcal O_R^*$, we note that $\overline{\mathcal O_L^2}$ and $\overline{\mathcal O_R^{*2}}$ are single-sided operators that are diagonal in the energy basis. Using \eqref{eq:ETHaverage} one has
\be\label{eq:diagop}
\bra{\epsilon_n}\overline{\mathcal O^2}\ket{\epsilon_m}
=
\sum_p \overline{\mathcal O_{np}\mathcal O_{mp}^*}
=
\delta_{nm}\sum_p \frac{g(\bar\epsilon_{np},\omega_{np})}{\varrho(\bar\epsilon_{np})},
\ee
so these terms can be absorbed into a shift of $H_0$ by shifting the energies
$\epsilon_n\mapsto \tilde\epsilon_n$. In the thermodynamic limit, the sum in \eqref{eq:diagop} is replaced by an integral, with additional factors of the density of states $\varrho(\epsilon)$, and we get
\be\label{eq:energyshift}
\delta \epsilon(\epsilon) := \tilde \epsilon(\epsilon)  - \epsilon = \dfrac{JN}{2}\int \text{d}\omega \dfrac{\varrho(\epsilon-\omega)}{\varrho(\epsilon- \frac{\omega}{2})}\,g\left(\epsilon- \frac{\omega}{2},\omega\right)\,.
\ee 

\paragraph{Cross kernel ---} The nontrivial left-right coupling comes from the averaged cross term. One has
\be
\langle \epsilon_n|\langle\epsilon_m^*|\,\mathcal O_L\mathcal O_R^*\,|\epsilon_p\rangle|\epsilon_q^*\rangle
=
\mathcal O_{np}\,\mathcal O_{mq}^*,
\ee
and therefore, using \eqref{eq:ETHform} and taking the average \eqref{eq:ETHaverage},
\be
\mathcal{K}_g:=\overline{\mathcal O_L\mathcal O_R^*}
=
\sum_{n,m}
\frac{g\!\left(\overline{\epsilon}_{nm},\omega_{nm}\right)}
{\varrho\!\left(\overline{\epsilon}_{nm}\right)}
\ket{\epsilon_n}\ket{\epsilon_n^*}\bra{\epsilon_m}\bra{\epsilon_m^*}.
\label{eq:KgOperator}
\ee
The kernel acts entirely within the diagonal subspace spanned by
$\{\ket{\epsilon_n}\ket{\epsilon_n^*}\}$ and annihilates off-diagonal states.

Up to the energy shift discussed above, at $K\to\infty$, by the law of large numbers over the operator ensemble, the Hamiltonian \eqref{eq:Hamiltonianinf} converges to
\be
H_\infty
=
H_{0,L}+H_{0,R}^* + (\overline{\mathcal O^2})_L + (\overline{\mathcal O^2})_R
- JN\,\mathcal{K}_g\,.
\label{eq:HinfETH}
\ee
We see that $H_\infty$ commutes with $H_{0,L}-H_{0,R}^*$, and it therefore preserves the diagonal subspace in this limit.

\subsection{ODE for ground state wavefunction}

Consider a smooth diagonal state $\ket{\psi} = \sum_{n} \varrho(\epsilon_n)^{-1/2}\,a_n\ket{\epsilon_n}\ket{\epsilon^*_n}$. To study the thermodynamic limit, it is convenient to work with a canonically normalized continuum wavefunction, where energy eigenstates $\ket{\epsilon}$ are delta-function normalized, $\bra{\epsilon}\ket{\epsilon'} = \delta(\epsilon-\epsilon')$. We therefore write the smooth diagonal state as
\be\label{eq:diagstate}
\ket{\psi}
=
\int \dd \epsilon \,a(\epsilon)\ket{\epsilon}\ket{\epsilon^*},
\qquad
\int \dd \epsilon\,|a(\epsilon)|^2=1\,.
\ee
This form is convenient because the wavefunction is normalized with the flat measure $\dd\epsilon$.

The cross kernel becomes
\be
 \mathcal{K}_g= \int \text{d}\epsilon \, \text{d}\epsilon' \sqrt{\varrho(\epsilon) \varrho(\epsilon')}\,\frac{g\!\left(\frac{\epsilon+\epsilon'}{2},\,\epsilon-\epsilon'\right)}
{\varrho\!\left(\frac{\epsilon+\epsilon'}{2}\right)}\ket{\epsilon}\ket{\epsilon^*}\bra{\epsilon'}\bra{\epsilon'^*}\,.
\label{eq:KdiagCont}
\ee

Using \eqref{eq:KdiagCont}, the eigenvalue equation $H\ket{\psi}=2\lambda\ket{\psi}$ becomes
\be
(\epsilon+\delta\epsilon-\lambda)\,a(\epsilon)
=
\frac{JN}{2}\int \dd \omega\;
g\!\left(\epsilon-\frac{\omega}{2},\omega\right)
\frac{\sqrt{\varrho(\epsilon)\varrho(\epsilon-\omega)}}{\varrho(\epsilon-\omega/2)}
\,a(\epsilon-\omega).
\label{eq:inteqphi}
\ee
Here we used $\omega=\epsilon-\epsilon'$ and $\bar\epsilon=\epsilon-\omega/2$.\\

\paragraph{Local derivative expansion ---}
We now assume that the wavefunction is supported on sufficiently small energy differences $\omega$, so that $a(\epsilon-\omega)$ may be expanded locally. The thermodynamic factors appearing in \eqref{eq:energyshift} and in \eqref{eq:inteqphi} can be expanded in the thermodynamic limit as
\begin{gather}
\frac{\varrho(\epsilon-\omega)}{\varrho(\epsilon-\omega/2)}=\exp\!\left[
S(\epsilon-\omega)-S(\epsilon-\omega/2)
\right] 
= e^{-\beta(\epsilon)\omega/2}\left(
1+O\!\left(\frac{\beta(\epsilon)^2\omega^2}{C(\epsilon)}\right)
\right)\,,\label{eq:ratio1}\\[.2cm]
\frac{\sqrt{\varrho(\epsilon)\varrho(\epsilon-\omega)}}{\varrho(\epsilon-\omega/2)}
 =
\exp\!\left[
\frac12 S(\epsilon)+\frac12 S(\epsilon-\omega)-S(\epsilon-\omega/2)
\right]
=1+O\!\left(\frac{\beta(\epsilon)^2\omega^2}{C(\epsilon)}\right)\,,
\label{eq:ratio2}
\end{gather}
for the inverse temperature and specific heat defined as
\be
\beta(\epsilon)=S'(\epsilon),\qquad
C(\epsilon)=\frac{\dd \epsilon}{\dd T}
=
-\frac{\beta(\epsilon)^2}{S''(\epsilon)}.
\ee
Since $\beta(\epsilon)\omega \ll \sqrt{C(\epsilon)}$ in the thermodynamic limit, we drop the subleading terms in the respective expansions.

Expanding the wavefunction as
\be
a(\epsilon-\omega)
=
a(\epsilon)-\omega a'(\epsilon)+\frac{\omega^2}{2}a''(\epsilon)+\dots,
\label{eq:phiTaylor}
\ee
and inserting \eqref{eq:ratio1} and \eqref{eq:ratio2} into \eqref{eq:inteqphi}, we obtain the ordinary differential equation
\be
-\frac{1}{2}\,M_2(\epsilon)\,a''(\epsilon)
+M_1(\epsilon)\,a'(\epsilon)
+\left[{2\epsilon}+V(\epsilon)\right]a(\epsilon)
\approx
{2\lambda}\,a(\epsilon),
\label{eq:ODEgroundstate}
\ee
where the coefficients are defined as
\be
 M_n(\epsilon)
:=JN
\int \dd \omega\;
g\!\left(\epsilon-\frac{\omega}{2},\omega\right)\omega^n\,,\qquad
V(\epsilon) := JN \int \dd \omega\;
g\!\left(\epsilon-\frac{\omega}{2},\omega\right)\left(e^{-\beta(\epsilon)\omega/2}-1\right)\,.
\label{eq:gmoments}
\ee
The expansion in small $\omega$ is justified if $a(\epsilon)$ is parametrically peaked in the thermodynamic limit, as will be the case for the ground state. An additional self-consistency condition is the Hermiticity of the differential operator \eqref{eq:ODEgroundstate} with respect to the diagonal inner product
$\langle \psi_1|\psi_2\rangle
=\int d\epsilon\,a_1^*(\epsilon)a_2(\epsilon)$, which requires
\be\label{eq:Hermiticity_condition}
M'_2(\epsilon)
=
-2\,M_1(\epsilon)\,.
\ee
This condition is approximately satisfied in the thermodynamic regime of interest, essentially by expanding $g(\epsilon-\frac{\omega}{2},\omega)\approx g(\epsilon,\omega) -\frac{\omega}{2} \,\partial_\epsilon g(\epsilon,\omega)$ in $M_1(\epsilon)$ and assuming that $g(\epsilon,\omega)$ is even in $\omega$.\\

\paragraph{Removal of the drift term ---}
The first-derivative term in \eqref{eq:ODEgroundstate} can be removed by the field redefinition
\be
a(\epsilon)=\dfrac{\psi(\epsilon)}{\sqrt{M_2(\epsilon)}}\,.
\label{eq:phipsi}
\ee
Substituting into \eqref{eq:ODEgroundstate}, we obtain
\be
-\frac{1}{2}M_2(\epsilon)\,\psi''(\epsilon)
+
V_{\rm eff}(\epsilon)\,\psi(\epsilon)
\approx
2\lambda\,\psi(\epsilon),
\label{eq:Schrpsi}
\ee
for the effective potential
\be
V_{\rm eff}(\epsilon)
=
2\epsilon
+V(\epsilon)
-\frac{1}{2}
\left[
\frac{M_1(\epsilon)^2}{M_2(\epsilon)}
+
M_1'(\epsilon)
\right]\approx 2\epsilon
+V(\epsilon),
\label{eq:Veffgeneral}
\ee
where in the last approximation we have neglected the two terms in parentheses, which are subleading in the thermodynamic limit.\\

\paragraph{Quadratic approximation ---}
We now expand the effective potential around its minimum $\epsilon=\epsilon_\star$, defined by $2+V'(\epsilon_\star)=0$, or explicitly,
\be
 \partial_{\epsilon_\star}\int \dd \omega\;
g\!\left(\epsilon_\star-\frac{\omega}{2},\omega\right)\left(1-e^{-\beta(\epsilon_\star)\omega/2}\right)
 = \frac{2}{JN}\,.
\ee

Expanding the potential to quadratic order,
\be
V_{\rm eff}(\epsilon)
\approx
V_{\rm eff}(\epsilon_\star)
+\frac12 V''(\epsilon_\star)(\epsilon-\epsilon_\star)^2,
\label{eq:Veffquad}
\ee
and freezing the kinetic coefficient as $M_2(\epsilon)\approx M_2(\epsilon_\star)$, Eq.~\eqref{eq:Schrpsi} becomes the harmonic oscillator equation
\be
-\frac12 M_2(\epsilon_\star)\psi''(\epsilon)
+
\frac12 V''(\epsilon_\star)(\epsilon-\epsilon_\star)^2\psi(\epsilon)
\approx
\big(2\lambda-V_{\rm eff}(\epsilon_\star)\big)\psi(\epsilon).
\label{eq:HOsch}
\ee
The normalized ground-state solution is therefore Gaussian,
\be
\psi_0(\epsilon)
=
\frac{1}{(\pi\sigma^2)^{1/4}}
\exp\!\left[
-\frac{(\epsilon-\epsilon_\star)^2}{2\sigma^2}
\right],
\label{eq:psigaussian}
\ee
with width
\be
\sigma^2
=
\sqrt{\frac{M_2(\epsilon_\star)}{V''(\epsilon_\star)}}\,.
\label{eq:sigma}
\ee
The spectrum of diagonal excitations on top of the ground state is gapped, with energies
\be\label{eq:2lambda} 
2\lambda_n = 2\epsilon_\star + V(\epsilon_\star) +\left(n+\frac12\right)\sqrt{M_2(\epsilon_\star)\,V''(\epsilon_\star)}. 
\ee 
Before evaluating $\sigma^2$ and $2\lambda_n$ more explicitly in specific systems, let us first note that $\sigma^2$ is the width parameter of the Gaussian wavefunction, so the corresponding energy variance is $\sigma^2/2$. In particular, $\sigma^2$ is extensive in $N$, and therefore so is the energy variance.

At this order, the field-redefinition factor is likewise subleading in the derivative expansion and may therefore be regarded as constant over the support of the wavefunction. Hence, we arrive at the Gaussian ground-state wavefunction
\be 
\ket{\rm GS} = \frac{1}{(\pi\sigma^2)^{1/4}}\int \text{d}\epsilon \exp\!\left(
-\frac{(\epsilon-\epsilon_\star)^2}{2\sigma^2}
\right)\ket{\epsilon}\ket{\epsilon^*}\,.
\ee 

\subsection{Overlap with TFD}

The TFD has a Gaussian wavefunction in the thermodynamic limit
\be 
\ket{\rm TFD} = \dfrac{1}{\sqrt{Z(\beta)}}\int \text{d}\epsilon \sqrt{\varrho(\epsilon)} e^{-\beta \epsilon/2} \ket{\epsilon}\ket{\epsilon^*}\approx \dfrac{1}{(\pi \sigma_\beta^2)^{1/4}} \int \text{d}\epsilon \exp\left(-\frac{(\epsilon-\epsilon_\beta)^2}{2\sigma_\beta^2}\right)\ket{\epsilon}\ket{\epsilon^*}\,,
\ee 
where the usual thermodynamic relations determine the canonical mean energy and variance,
\be 
S'(\epsilon_\beta) = \beta\,,\qquad
(\Delta H_0)^2_\beta := \langle H_0^2\rangle_\beta-\langle H_0\rangle_\beta^2 = \dfrac{C(\epsilon_\beta)}{\beta^2}\,.
\ee 
Since $\sigma_\beta^2$ is the width parameter of the wavefunction rather than the variance of $|a(\epsilon)|^2$, the two are related by
\be
\sigma_\beta^2 = 2\,(\Delta H_0)^2_\beta=\frac{2\,C(\epsilon_\beta)}{\beta^2}\,.
\ee

The overlap with the ground state is then
\be
|\bra{\rm TFD}\ket{\rm GS}| =
\sqrt{\dfrac{2\sigma \sigma_\beta}{\sigma^2 + \sigma_\beta^2}}
\exp \left(-\dfrac{(\epsilon_\star-\epsilon_\beta)^2}{2(\sigma^2 + \sigma_\beta^2)}\right) \,.
\ee

The maximal overlap is attained by tuning $\beta = \beta_\star := \beta(\epsilon_\star)$ such that $\epsilon_{\beta_\star} = \epsilon_\star$ and the peaks of the Gaussians coincide. The corresponding fidelity is
\be\label{eq:fidelitymaxtheory}
\mathcal{F} = \max_{\beta} |\bra{\rm TFD}\ket{\rm GS}|^2 =
{\dfrac{2\sigma \sigma_{\beta_\star}}{\sigma^2 + \sigma_{\beta_\star}^2}} \,,
\ee
which remains finite in the thermodynamic limit, as both $\sigma^2$ and $\sigma^2_{\beta_\star}$ are extensive.

\subsection{Physical examples}

We now consider some specific examples where we can evaluate the width parameter $\sigma^2$, the corresponding fidelity $\mathcal{F}$, and the energy gap $\Delta:=2\lambda_1-2\lambda_0$ as a function of the coupling $J$ of the $LR$ interaction. 

It is convenient to note that in the thermodynamic limit one has
\be 
V(\epsilon) = \dfrac{JN}{2}\bra{\rm TFD_{\beta}} \left(\mathcal{O}_{L}- \mathcal{O}^*_{R}\right)^2\ket{\rm TFD_{\beta}}\big|_{\beta = \beta(\epsilon)} =JN\left[ G_{\beta}(0) - G_{\beta}(\beta/2)\right]\big|_{\beta = \beta(\epsilon)}\,,
\ee 
for the Euclidean thermal two-point function
\be
G_{\beta}(\tau)
=
\frac{1}{Z(\beta)}
\Tr\!\left(e^{-\beta H_0}\,\mathcal O(\tau)\mathcal O(0)\right).
\ee

{\bf Lattice systems at high temperatures.} For lattice systems, a simple estimate is obtained in the high-temperature regime. Regularity of the Euclidean correlator at $\beta\to0$ gives
\be
G_\beta(0)-G_\beta(\beta/2)
=
\frac{\beta^2}{8}\,\mu_2+O(\beta^4),
\qquad
\mu_2:=\int d\omega\, g(\epsilon_\infty,\omega)\,\omega^2 .
\ee
On the other hand, expanding the energy around infinite temperature in terms of the infinite-temperature energy variance,
\be
\epsilon(\beta)=\epsilon_\infty-(\Delta H_0)^2_{\beta=0}\,\beta+O(\beta^2),
\qquad
(\Delta H_0)^2_{\beta=0}
=
-\left.\frac{d\epsilon}{d\beta}\right|_{\beta=0}
=
\langle H_0^2\rangle_{\beta =0}-\langle H_0\rangle_{\beta =0}^2 ,
\ee
one obtains
\be
V(\epsilon)
=
\frac{JN\,\mu_2}{8\big((\Delta H_0)^2_{\beta=0}\big)^2}\,(\epsilon_\infty-\epsilon)^2+\dots
\quad \Rightarrow\quad
V_{\rm eff}(\epsilon)
=
2\epsilon+
\frac{JN\,\mu_2}{8\big((\Delta H_0)^2_{\beta=0}\big)^2}\,(\epsilon_\infty-\epsilon)^2+\dots .
\ee
Hence, the saddle-point energy lies at
\be
\epsilon_\star
=
\epsilon_\infty-\frac{8\big((\Delta H_0)^2_{\beta=0}\big)^2}{JN\,\mu_2},
\ee
and the associated inverse temperature is non-extensive and inversely proportional to the $LR$ coupling,
\be \label{eq:betalattice}
\beta_\star
=
\frac{8(\Delta H_0)^2_{\beta=0}}{JN\,\mu_2}.
\ee 
Moreover,
\be
V''(\epsilon_\star)\simeq \frac{JN\,\mu_2}{4\big((\Delta H_0)^2_{\beta=0}\big)^2},
\qquad
M_2(\epsilon_\star)\simeq JN\,\mu_2 ,
\ee
and therefore the Gaussian width \eqref{eq:sigma} and oscillator gap from \eqref{eq:2lambda} are
\begin{gather}
\sigma^2
=
\sqrt{\frac{M_2(\epsilon_\star)}{V''(\epsilon_\star)}}
=
2(\Delta H_0)^2_{\beta=0},\label{eq:sigmalattice}
\\[.2cm]
\Delta=
\sqrt{M_2(\epsilon_\star)V''(\epsilon_\star)}
=
\frac{JN\,\mu_2}{2(\Delta H_0)^2_{\beta=0}}.
\end{gather}
In particular, the width parameter $\sigma^2$ is extensive in $N$, so the corresponding ground-state energy variance, equal to $\sigma^2/2$, is also extensive. By contrast, the gap remains non-extensive, $\Delta\sim O(N^0)$, in the thermodynamic limit.\\

\paragraph{Fidelity ---} For the TFD one has
\be
(\Delta H_0)^2_\beta=\frac{C(\beta)}{\beta^2},
\ee
and therefore the width parameter of the TFD wavefunction is
\be
\sigma_\beta^2 = 2(\Delta H_0)^2_\beta
=2\,\frac{C(\beta)}{\beta^2}.
\ee
Using the high-temperature expansion above, it follows that
\be
C(\beta)=\beta^2(\Delta H_0)^2_{\beta=0}+O(\beta^3),
\qquad
\sigma_{\beta_\star}^2\simeq 2(\Delta H_0)^2_{\beta=0},
\ee
where $\beta_\star = \beta(\epsilon_\star)$ is the optimal inverse temperature. Therefore, using \eqref{eq:sigmalattice}, we find
\be
\sigma^2=\sigma_{\beta_\star}^2,
\ee
so the two Gaussian wavefunctions coincide and the fidelity \eqref{eq:fidelitymaxtheory} gives
\be
\mathcal F = 1.
\ee
Thus, already within the Gaussian approximation, the ground state and the TFD coincide in this high-temperature regime.

{\bf Quantum critical systems.}
For a quantum critical system in $d$ spatial dimensions, the scale-invariant equation of state takes the form
\be
\epsilon(\beta)\sim N\,\beta^{-(d+1)},
\qquad
C(\beta)\sim N\,\beta^{-d},
\ee
so that
\be
\beta(\epsilon)\sim \left(\frac{N}{\epsilon}\right)^{\frac{1}{d+1}}.
\ee
For an operator $\mathcal O$ of scaling dimension $\Delta_{\mathcal{O}}$, we approximate
\be
G_\beta(0)-G_\beta(\beta/2)\sim -c_{\mathcal O}\,\beta^{-2\Delta_{\mathcal{O}}},
\qquad c_{\mathcal O}>0,
\ee
after whatever short-distance regularization or smearing is implicit in the operator entering the parent Hamiltonian. It then follows that
\be
V(\epsilon)
=
JN\big[G_\beta(0)-G_\beta(\beta/2)\big]_{\beta=\beta(\epsilon)}
\sim
-J\,c_{\mathcal O}\,N^{1-\alpha}\epsilon^\alpha,
\qquad
\alpha:=\frac{2\Delta_{\mathcal{O}}}{d+1}.
\ee
Hence
\be
V_{\rm eff}(\epsilon)\approx 2\epsilon+V(\epsilon)\sim 2\epsilon-J\,c_{\mathcal O}\,N^{1-\alpha}\epsilon^\alpha.
\ee
For $\alpha<1$, namely $2\Delta_{\mathcal{O}}<d+1$, this potential develops a minimum at
\be
\epsilon_\star\sim N\,J^{\frac{1}{1-\alpha}}
=
N\,J^{\frac{d+1}{d+1-2\Delta_{\mathcal{O}}}},
\qquad
\beta_\star\sim J^{-\frac{1}{d+1-2\Delta_{\mathcal{O}}}},
\ee
We note that the Maldacena-Qi ground state \cite{maldaqi} follows this scaling for the inverse temperature (for $d=1$) because of the emergent conformal symmetry that describes the infrared of the SYK model.

The curvature is
\be
V_{\rm eff}''(\epsilon_\star)\sim
N^{-1}J^{-\frac{d+1}{d+1-2\Delta_{\mathcal{O}}}}.
\ee
Using the general formulas \eqref{eq:sigma}, \eqref{eq:2lambda}, and \eqref{eq:fidelitymaxtheory}, together with the dimensional estimate
\be
M_2(\epsilon_\star)\sim JN\,\beta_\star^{-(2\Delta_{\mathcal{O}}+2)},
\ee
we obtain
\be\label{eq:gapconformalgen}
\sigma^2\sim N\,J^{\frac{d+2}{d+1-2\Delta_{\mathcal{O}}}},
\qquad
\Delta\sim J^{\frac{\Delta_{\mathcal{O}}+1}{d+1-2\Delta_{\mathcal{O}}}}.
\ee
For the Maldacena-Qi model \cite{maldaqi}, the gap in \eqref{eq:gapconformalgen} is slightly larger than the true gap because the fundamental excitations are not totally diagonal, as they come from matter excitations in the wormhole.

\paragraph{Fidelity ---} Moreover, the width parameter of the TFD wavefunction follows the thermal variance as
\be
\sigma_{\beta_\star}^2=2\,\frac{C(\beta_\star)}{\beta_\star^2}
\sim
N\,J^{\frac{d+2}{d+1-2\Delta_{\mathcal{O}}}},
\ee
so the maximal fidelity \eqref{eq:fidelitymaxtheory} is
\be
\mathcal F=
{\frac{2\sigma\sigma_{\beta_\star}}{\sigma^2+\sigma_{\beta_\star}^2}}\sim O(N^0J^0)\,.
\ee 
Thus, within this approximation, the maximal fidelity remains $O(1)$ in the thermodynamic limit, with no leading dependence on $J$.

\subsection{Fidelity loss at finite $K$}

The analysis above determines the ground state for $K=\infty$ because the Hamiltonian converges, by the law of large numbers, to the operator-averaged Hamiltonian $H_\infty$, where it preserves the diagonal subspace spanned by $\{\ket{\epsilon_n}\ket{\epsilon_n^*}\}$. At finite $K$, leakage out of the diagonal subspace is generated by fluctuations of all three pieces in $(\mathcal O_L-\mathcal O_R^*)^2
=
\mathcal O_L^2+\mathcal O_R^{*2}-2\mathcal O_L\mathcal O_R^*$. For example, the fluctuating part of the cross kernel may be modeled using ETH as
\be
\delta \mathcal K_g
=
\frac{1}{\sqrt K}
\sum_{n,m,p,q}
\frac{g\!\left(\overline{\epsilon}_{nm},\omega_{nm}\right)^{1/2}
      g\!\left(\overline{\epsilon}_{pq},\omega_{pq}\right)^{1/2}}
{\varrho\!\left(\overline{\epsilon}_{nm}\right)^{1/2}
 \varrho\!\left(\overline{\epsilon}_{pq}\right)^{1/2}}
\,X_{nmpq}\,
\ket{\epsilon_n}\ket{\epsilon_m^*}\bra{\epsilon_p}\bra{\epsilon_q^*},
\label{eq:deltaKg}
\ee
where $X_{nmpq}$ is a centered random tensor with order-one variance. Note the overall $K^{-1/2}$ suppression, which follows from the central-limit scaling of the sum over $K$ independent operators.

Let $\ket{\psi_\infty}$ denote the Gaussian ground state of $H_\infty$ obtained in the previous subsection, with optimal TFD fidelity $\mathcal F_\infty$ given by \eqref{eq:fidelitymaxtheory}. Let also
\be
P_{\rm diag}
=
\sum_n
\ket{\epsilon_n}\ket{\epsilon_n^*}\bra{\epsilon_n}\bra{\epsilon_n^*},
\qquad
P_{\rm off}=1-P_{\rm diag},
\ee
be the projectors onto the diagonal and off-diagonal subspaces. Since the TFD lies entirely in the diagonal subspace, any finite-$K$ admixture into the off-diagonal sector necessarily lowers the fidelity with the TFD.

A first perturbative estimate gives the off-diagonal probability leakage
\be
\|P_{\rm off}\ket{{\rm GS}}\|^2
\;\lesssim\;
\frac{\|P_{\rm off}\,\delta H\,\ket{\psi_\infty}\|^2}{\Delta_{\rm off}^2},
\label{eq:offdiagleakageThermal}
\ee
where we have defined the off-diagonal gap
\be
\Delta_{\rm off}:=
E_{\rm off}^{\rm min} - 2\lambda_0^{(\infty)},
\ee
namely, the gap between the diagonal ground-state energy of $H_\infty$ and the bottom of the off-diagonal sector. Since $\delta H$ is a sum of $K$ independent centered terms with overall prefactor $JN/K$, one finds
\be
\overline{\|P_{\rm off}\,\delta H\,\ket{\psi_\infty}\|^2}
\;\sim\;
\frac{(JN)^2}{K},
\ee
up to order-one factors controlled by thermal correlators at $\beta_\star$. Hence
\be
\overline{\|P_{\rm off}\ket{{\rm GS}}\|^2}
\;\sim\;
\frac{(JN)^2}{K\,\Delta_{\rm off}^2}.
\label{eq:offdiagweightThermal}
\ee

Since the TFD lies entirely in the diagonal subspace, the overlap with the exact finite-$K$ ground state is bounded by the diagonal weight:
\be
|\braket{{\rm TFD}_{\beta_\star}}{{\rm GS}}|^2
\le
\|P_{\rm diag}\ket{{\rm GS}}\|^2
=
1-\|P_{\rm off}\ket{{\rm GS}}\|^2.
\label{eq:fidelityboundoffdiag}
\ee

This estimate is sufficient to show that the diagonal analysis above is perturbatively reliable only when
\be
K\gg \left(\frac{JN}{\Delta_{\rm off}}\right)^2.
\label{eq:Kconditionthermal}
\ee

What should we expect for the off-diagonal gap $\Delta_{\rm off}$? On general grounds, it is natural to expect it to be of the same order as the diagonal gap, which remains non-extensive in the thermodynamic limit. This is also what is seen in studies of the interaction term in the parent Hamiltonian, that is, the strong-coupling regime of the parent Hamiltonian, for both SYK and spin models at finite $K$ and $N$, such as \cite{Jian:2022pvj,Guo:2024zmr}. In those cases, the gap is set by the degree of nonlocality of the fermion or spin terms in the Hamiltonian. From the ETH-based discussion alone, however, this is not obvious, since ETH effectively models the operators as fully nonlocal random matrices, at least within the Gaussian ETH approximation.

Assuming an off-diagonal gap of order $O(N^0)$, \eqref{eq:Kconditionthermal} implies that finite-$K$ leakage can remain large unless $K \gtrsim O(N^2)$. In the spin chain we study, where $K=2N$, this condition is not met, and the ground state is therefore not perturbatively close to the diagonal subspace. This matches our numerical results, which show a clear exponential decay of fidelity in $N$. More generally, the fidelity argument of \cite{Cottrell_2019} is not controlled unless \eqref{eq:Kconditionthermal} is satisfied, in which case one needs some further suppression of the leakage into the off-diagonal sector. As discussed in the main text, the penalty term \eqref{eq:modified_Hamiltonian_TFD} provides exactly this by lifting the off-diagonal sector and restoring perturbative control \eqref{eq:offdiagweightpenalty}.
\section{Ground state in random matrix theory}
\label{sec:rmt}

In this appendix, we develop a random matrix theory (RMT) version of the coupled Hamiltonian \eqref{eq:Hamiltonianprop} and study its ground state analytically. We work on a finite-dimensional Hilbert space of dimension $L$ (where $L=2^N$ for the case of $N$ qubits).  Within this Hilbert space, we construct the Hamiltonian by selecting the Hermitian operators $\{\mathcal{O}_{\alpha}\}_{\alpha=1}^K$ independently from some distribution of random operators (with respect to the eigenbasis of $H_0$), with the two moments,
\be
\overline{\mathcal{O}_{nm}}=0\,,\qquad \overline{\mathcal{O}_{nm}\mathcal{O}_{pq}^*}= \dfrac{1}{L}\delta_{np}\delta_{mq}\,.
\ee

By the central limit theorem, in the limit of a large number of operators $K\to \infty$, with unit probability, the Hamiltonian \eqref{eq:Hamiltonianprop} becomes
\be\label{eq:Hamiltonianinf} 
H_\infty =   H_{0,L} + H^\ast_{0,R} + \dfrac{JN}{2}\overline{\left(\mathcal{O}_L-\mathcal{O}_R^*\right)^2} = H_{0,L} + H^\ast_{0,R} +  JN \left(\mathbf{1}-\ket{\infty}\bra{\infty}\right)\,,
\ee 
where $\ket{\infty}$ is the infinite temperature TFD state
\be 
\ket{\infty} = \dfrac{1}{\sqrt{L}}\sum_{n=1}^L \ket{\epsilon_n}\ket{\epsilon_n^*}\,.
\ee

\subsection{Exact diagonalization}

The off-diagonal states $|\epsilon_n\rangle \ket{\epsilon^*_m}$ are eigenvectors of $H_\infty$ with energy $\epsilon_n + \epsilon_m + JN$. In the diagonal subspace, we write an eigenfunction as the Choi state of the diagonal operator
\be 
\ket{A} = \sum_{n=1}^L A_{n}\ket{\epsilon_n}\ket{\epsilon_n^*}\qquad \Leftrightarrow\qquad A = \sum_{n=1}^L A_{n} \ket{\epsilon_n}\bra{\epsilon_n}\,.
\ee 
The diagonal eigenstates $H_\infty\ket{A} = \left(2\lambda + JN \right)\ket{A}$ satisfy
\be\label{eq:rmteigenv} 
(H_0-\lambda) A = \dfrac{JN}{2L} \text{Tr}(A)\;\Rightarrow\; \dfrac{A}{\text{Tr}(A)} = \dfrac{JN}{2L} R(\lambda)\,,
\ee 
where we have introduced the resolvent matrix of the bare Hamiltonian
\be 
R(\lambda) = \dfrac{1}{H_0-\lambda}\,.
\ee 
Taking the trace of \eqref{eq:rmteigenv} yields the eigenvalue equation
\be\label{eq:lambdarmt} 
\text{Tr}\,R(\lambda) = \dfrac{2L}{JN}\,.
\ee 
So, we obtain that the ground state is the Choi state of the resolvent $R(\lambda)$ of the bare Hamiltonian $H_0$, for some coupling-dependent value of $\lambda$, which we denote $\lambda = \lambda_\star$:
\be 
|{\rm GS}\rangle = \dfrac{1}{\sqrt{\text{Tr}\left(R(\lambda_\star)^2\right)}}\,|R(\lambda_\star)\rangle\,=\, \dfrac{1}{\sqrt{\sum_{m}(\epsilon_m-\lambda_\star)^{-2}}} \sum_{n=1}^L\dfrac{1}{\epsilon_n - \lambda_\star}|\epsilon_n\rangle |\epsilon^*_n\rangle\,.
\ee 
Note that not only the ground state, but all of the diagonal eigenstates of $H_\infty$ have resolvent Choi form $|R(\lambda)\rangle$, with different $\lambda$. The value of $\lambda_\star$ is the smallest solution to \eqref{eq:lambdarmt}.

\paragraph{Ground state energy ---} Let us solve for $\lambda_\star$. We rewrite \eqref{eq:lambdarmt} in the form
\be\label{eq:stieltjes}
\int \text{d}\epsilon\, \hat{\varrho}(\epsilon) \dfrac{1}{\epsilon-\lambda} = \dfrac{2}{JN}
\ee 
where we defined the (unit-normalized) density of states of $H_0$,
\be 
\hat{\varrho}(\epsilon)  := \dfrac{1}{L}\sum_{n=1}^L \delta(\epsilon- \epsilon_n)\,,\qquad \,\int \text{d}\epsilon\, \hat{\varrho}(\epsilon) = 1\,.
\ee 
The equation to solve \eqref{eq:stieltjes} is the standard Stieltjes transform inversion. We define the Stieltjes transform of the density of states
\be
m(\lambda)
:=
\int \text{d}\epsilon\,\hat{\varrho}(\epsilon)\,
\frac{1}{\epsilon-\lambda}.
\ee
Equation~\eqref{eq:stieltjes} then reads
\be\label{eq:inversestieltjes}
m(\lambda)=\frac{2}{JN}.
\ee

For a discrete spectrum $\text{supp}(\hat{\varrho}) \subset [\epsilon_{\min},\epsilon_{\max}]$, there are $L$ solutions to this equation for fixed $JN$ (one in between each pair of neighboring eigenvalues of $H_0$). These correspond to all the diagonal eigenvalues of $H_\infty$.
For $\lambda<\epsilon_{\min}$,
\be
m'(\lambda)
=
\int \text{d}\epsilon\,\hat{\varrho}(\epsilon)\,
\frac{1}{(\epsilon-\lambda)^2}
>0,
\ee
so $m(\lambda)$ is strictly increasing on $(-\infty,\epsilon_{\min})$.
Its limiting behavior is
\be
\lim_{\lambda\to -\infty} m(\lambda)=0^+,
\qquad
\lim_{\lambda\to \epsilon_{\min}^-} m(\lambda)=+\infty.
\ee
Therefore, for any $JN>0$, there exists a unique solution 
\be\label{eq:coniditongsermt}
\lambda_\star<\epsilon_{\min}\,,
\ee
to Eq.~\eqref{eq:inversestieltjes}. Recall that the ground state energy is then $2\lambda_\star + JN$. By virtue of \eqref{eq:coniditongsermt}, this is always smaller than the energy for the off-diagonal eigenstates.\\

\paragraph{Gap ---} The first excited states are again diagonal states, corresponding to the next solution of \eqref{eq:lambdarmt}, for which $\lambda$ lies between the two lowest eigenvalues of $H_0$. It follows that $H_\infty$ is gapped, with
\be
\Delta \ge \Delta_0:=2(\epsilon_{\min}-\lambda_\star)\,.
\ee
If the low-lying spectrum is dense, the lower bound will be close to being saturated.

\subsection{Fidelity with a finite temperature TFD}

The overlap probability between the ground state $\ket{\rm GS}$ and the TFD is
\begin{gather}\label{eq:RMTfidelity}
\left|\bra{ {\rm TFD}_\beta}\ket{\rm GS}\right|^2 = \dfrac{\left|\text{Tr}\left(e^{-\beta H_0/2}R(\lambda_\star)\right)\right|^2}{\text{Tr}\left(e^{-\beta H_0}\right)\text{Tr}\left(R(\lambda_\star)^2\right)}
=
\frac{
\left[
\int \text{d}\epsilon\, \hat{\varrho}(\epsilon)\,
\frac{e^{-\beta \epsilon/2}}{\epsilon-\lambda_\star}
\right]^2
}{
\left[
\int \text{d}\epsilon\, \hat{\varrho}(\epsilon)\,
\frac{1}{(\epsilon-\lambda_\star)^2}
\right]
\left[
\int \text{d}\epsilon\, \hat{\varrho}(\epsilon)\,
e^{-\beta \epsilon}
\right]
}\,.
\end{gather}
This quantity remains finite as $N\to \infty$ (or equivalently as $L= 2^N \to \infty$). For example, using \eqref{eq:stieltjes} and that $\text{Tr}\,(R(\lambda_\star)^2) \leq L(\epsilon_{\min}-\lambda_\star)^2$, the overlap probability with the infinite temperature TFD is lower bounded by
\be 
\left|\bra{ \infty}\ket{\rm GS}\right|\geq \dfrac{\Delta_0}{JN}\,,
\ee 
so it is set by the gap $\Delta_0$, which for large enough $J$ becomes extensive.

\subsection{Example: GUE Hamiltonian}

Up to this point, the analysis holds for an individual $H_0$ with a discrete spectrum. We now coarse-grain the spectrum of $H_0$, thinking of it as a draw from an ensemble of random Hamiltonians. For simplicity, we consider a Hamiltonian drawn from the GUE. The unit-normalized large-$N$ density of states is the semicircle density
\be\label{eq:semicircle}
\hat{\varrho}(\epsilon)
=
\frac{1}{2\pi\sigma^2}
\sqrt{4\sigma^2-\epsilon^2}\,
\Theta(2\sigma-|\epsilon|)\,.
\ee
This model is good for $H_0$ if we focus on the physics in the ``bulk'', away from the edge of the spectrum (for instance, if $JN\to 0$, the continuous density of states will not capture properties of the ground state of the coupled Hamiltonian correctly; one must use the discrete spectrum instead).

In this case, the Stieltjes transform is defined away from the branch cut 
\be
m(\lambda)
=
\int \text{d}\epsilon\,
\frac{\hat{\varrho}(\epsilon)}{\epsilon-\lambda},
\qquad
\lambda\notin[-2\sigma,2\sigma].
\ee
Using \eqref{eq:semicircle}, one gets
\be
m(\lambda)
=
\frac{\sqrt{\lambda^2-4\sigma^2}-\lambda}{2\sigma^2},
\ee
where the branch cut is at $|\lambda|<2\sigma$.\\

\paragraph{Critical coupling ---}
Since
\be
\lim_{\lambda\to -2\sigma^-} m(\lambda)=\frac{1}{\sigma},
\ee
the eigenvalue equation \eqref{eq:inversestieltjes} admits a real solution only if
\be
J \ge J_c := \dfrac{2\sigma}{N}.
\ee
Thus, the semicircle continuum model does not describe the
$J\to 0$ regime; that limit requires a discrete spectrum.

For $J\ge J_c$, the solution for the ground state eigenvalue and the gap is
\be\label{eq:gapGUE}
\lambda_\star
=
-\dfrac{JN}{2} \left(1+\dfrac{J^2_c}{J^2}\right)\quad\Rightarrow \quad\Delta_0 = 2(-2\sigma - \lambda_\star) = JN \left(1-\dfrac{J^2_c}{J^2}\right)\,.
\ee

\paragraph{Fidelity ---} We will now evaluate the fidelity. The normalization of the ground state is
\be
\int \mathrm{d}\epsilon\,\hat{\varrho}(\epsilon)\,
(\epsilon-\lambda_\star)^{-2}
=
-\frac{1}{2\sigma^2}
-\frac{\lambda_\star}{2\sigma^2\sqrt{\lambda_\star^2-4\sigma^2}}
=
\frac{4}{(JN)^2}
\left(1-\dfrac{J^2_c}{J^2}\right)^{-1}.
\ee
and the canonical partition function is
\be
\int \text{d}\epsilon\,
\varrho(\epsilon)
e^{-\beta\epsilon}
=
\frac{I_1(2\sigma\beta)}{\sigma\beta}.
\ee
where $I_\nu(x)
=
\sum_{m=0}^\infty
\frac{1}{m!\,\Gamma(m+\nu+1)}
\left(\frac{x}{2}\right)^{2m+\nu}$ is the modified Bessel function of the first kind.

\begin{figure}
    \centering
    \includegraphics[width=0.7\linewidth]{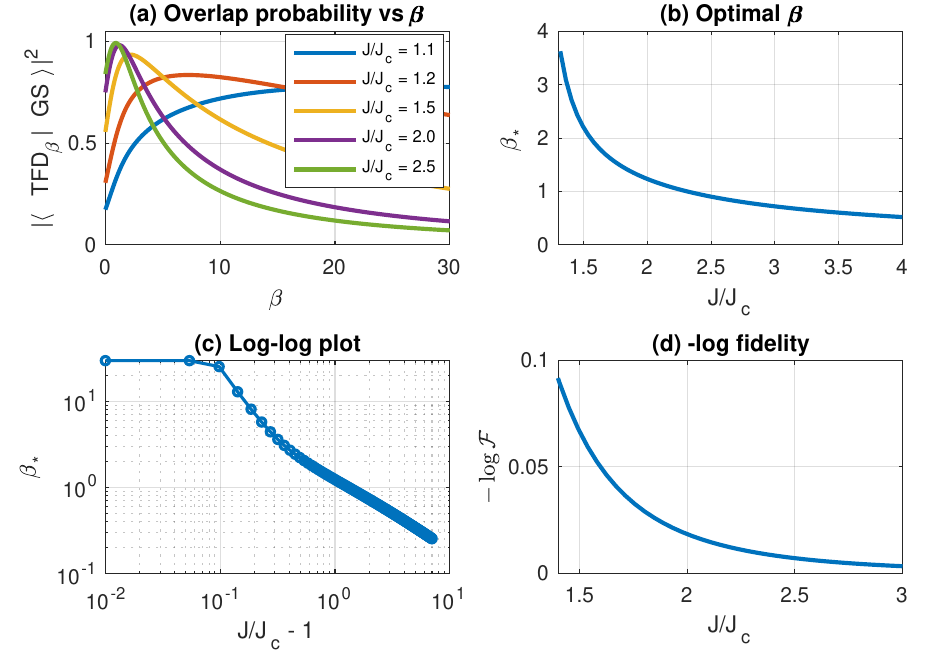}
    \caption{
Semicircle overlap diagnostics. 
(a) Overlap probability $|\langle {\rm TFD}_\beta | {\rm GS} \rangle|^2$ as a function of $\beta$ for several values of $J/J_c$. 
(b) Value $\beta_\star$ (in units of $1/\sigma$) that maximizes the overlap, shown as a function of $J/J_c$. 
(c) Log-log plot of $\beta_\star$ versus $J/J_c-1$ 
(d) Log fidelity $-\log \mathcal{F}$ as a function of $J/J_c$. Near $J/J_c\simeq 1$, the RMT description of $H_0$ in terms of the semicircle density of states becomes unreliable, since the overlap is then sensitive to edge effects and the discrete low-lying spectrum.
}
    \label{fig:fidewigner}
\end{figure}

The numerator in \eqref{eq:RMTfidelity} can be expressed in terms of the Bessel series
\be
\int d\epsilon\,\hat\varrho(\epsilon)\,
\frac{e^{-\beta\epsilon/2}}{\epsilon-\lambda_\star}
=
\frac{2}{\sigma^2 \beta}
\sum_{m=1}^\infty
m\,\left(\dfrac{J_c}{J}\right)^m I_m(\sigma \beta)\,.
\ee
Then, the overlap probability with the TFD is
\be 
\left|\bra{ {\rm TFD}_\beta}\ket{\rm GS}\right|^2 = \dfrac{4}{\sigma \beta}\left(\frac{J^2}{J^2_c}-1\right)\dfrac{\left(\sum_{m=1}^\infty m (J_c/J)^m I_{m}(\sigma\beta)\right)^2}{I_1(2\sigma \beta)}\,.
\ee 
We plot the overlap probability in Fig. \ref{fig:fidewigner} for different values of $J/J_c\geq 1$. We find that there is a coupling-dependent maximum $\beta_\star(J/J_c)$ that maximizes the overlap. We plot the fidelity $\mathcal{F}=\max_\beta |\langle {\rm TFD}_\beta | {\rm GS} \rangle|^2$ as a function of the coupling. These results show that the fidelity attains an $O(1)$ value in the large-$N$ limit.\\

\paragraph{Strong coupling ---} At large coupling, $J/J_c \gg 1$, we can further expand the exact expression at small $\sigma\beta$ to obtain
\be
\left|\bra{ {\rm TFD}_\beta}\ket{\rm GS}\right|^2
\approx 
\left(1-\frac{J_c^2}{J^2}\right)
\left(
1+\frac{\sigma \beta J_c}{J}-\frac{(\sigma \beta)^2}{4}\right)\,.
\ee
Maximizing with respect to $\beta$ then gives
\be
\beta_\star \simeq \frac{2J_c}{\sigma J}.
\ee
This behavior is shown in Fig.~\ref{fig:fidewigner}, where the log-log slope of $\beta_\star$ as a function of $J/J_c$ approaches $-1$.

\subsection{Finite-$K$ leakage into the off-diagonal subspace}

Let $H=H_{\infty}+\delta H$, where $H_{\infty}$ is the $K\to\infty$ averaged Hamiltonian \eqref{eq:Hamiltonianinf} and $\delta H$ is the finite-$K$ fluctuation. 
If $\Delta_0 = 2(\epsilon_{\min} -\lambda_\star)$, standard perturbation theory gives
\be
\|P_{\rm off}|{\rm GS}\rangle\|^2
\;\lesssim\;
\frac{\|P_{\rm off}\delta H|\psi_\infty\rangle\|^2}{\Delta_0^2},
\ee
where $|\psi_\infty\rangle$ is the ground state of $H_{\infty}$ and $P_{\rm off} = \sum_{n\neq m}\ket{\epsilon_n}\ket{\epsilon_m^*}\bra{\epsilon_n}\bra{\epsilon_m^*}$ is the orthogonal projector onto the off-diagonal subspace.  
Since $\delta H$ is a sum of $K$ independent centered random terms with an overall prefactor $JN/K$, its typical size is reduced by the central-limit factor $K^{-1/2}$.  That is,
\be
\overline{\|P_{\rm off}\delta H|\psi_0\rangle\|^2}
\;\sim\;
\frac{(JN)^2}{K},
\ee
and hence
\be
\overline{\|P_{\rm off}|{\rm GS}\rangle\|^2}
\;\sim\;
\frac{(JN)^2}{K\Delta_0^2}\,.
\ee
up to constants of order one. Thus, the off-diagonal weight of the ground state is perturbatively small only when $K\gg (JN/\Delta_0)^2$. In this regime, the overlap with the TFD remains $O(N^0)$ in the large-$N$ limit, mainly because the finite $K$ ground state overlaps substantially with the $K\to \infty$ ground state.

\begin{figure}[t]
    \centering
    \includegraphics[width=.7\linewidth]{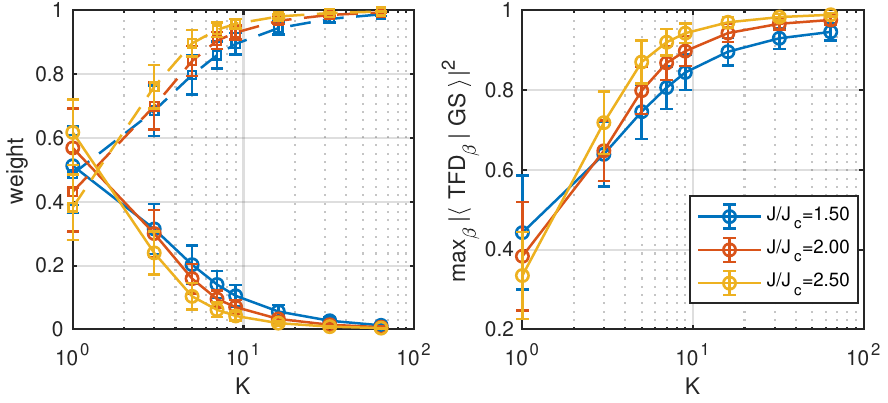}
    \caption{
Finite-$K$ and small $N$ numerics for the RMT coupled Hamiltonian. 
Left: off-diagonal weight $w_{\rm off}:= \|P_{\rm off}|{\rm GS}\rangle\|^2$ (solid) and diagonal weight $1-w_{\rm off}$ (dashed) of the ground state as functions of $K$. 
Right: optimal finite temperature TFD fidelity, $\max_{\beta}|\langle {\rm TFD}_{\beta}|{\rm GS}\rangle|^2$, as a function of $K$. Error bars denote the standard deviation over disorder realizations of the choice of $K$ Hermitian operators, using $100$ realizations, for $N=5$ qubits ($L=32$). The colors in both plots represent the values of the coupling in the legend.
}
    \label{fig:finiteK_diagnostics}
\end{figure}

For the RMT models, one finds that this regime is attained with a few operators $K\gtrsim O(N^0)$ because the gap $\Delta_0$ is extensive. This can be traced back to the fact that the system is completely non-local (as opposed to the non-extensive gap in the mixed field Ising or SYK examples discussed throughout the paper). For example, for the GUE Hamiltonian, using \eqref{eq:gapGUE}, we get
\be
 \overline{\|P_{\rm off}|{\rm GS}\rangle\|^2}
\;\sim\; \dfrac{1}{K} \left(1 - \dfrac{J^2_c}{J^2}\right)^{-2}\,.
\ee 
 Thus, the large overlap just requires $K \gg \left(1 - {J^2_c}/{J^2}\right)^{-2}$, which remains $O(N^0)$ in the thermodynamic limit if $1 - J_c/J = O(N^0)$. For $J\gg J_c$, this just tells us that $K\gtrsim  1$. This prediction is broadly consistent with what we find numerically for finite $K$ and small $N$, as shown in Fig. \ref{fig:finiteK_diagnostics}.

\section{Details of the Maldacena-Qi wormhole calculations}
\label{sec:SYK}

In this appendix, we consider two $q$-body SYK models on the left ($L$) and right ($R$) defined by the bare Hamiltonians
\be
  H_{{\rm SYK},L/R}=\mathrm{i}^{\frac{q}{2}}\sum_{i_1<\cdots < i_q} J_{i_1\cdots i_q} \psi_{i_1,L/R} \dots \psi_{i_q,L/R},\qquad \langle J^2_{i_1...i_q}\rangle = \dfrac{\mathcal{J}^2N}{2q^2{N \choose q}}\,.
\ee
Each system contains $N$ Majorana fermions $\psi_{i,L/R}$ ($i=1,\dots,N$), with normalization $\{\psi_i,\psi_j\}=2\delta_{ij}$. We couple the two SYK systems through the Maldacena-Qi interaction \cite{maldaqi}
\be\label{eq:intMQ}
  H_{\mathrm{int}}= \mathrm{i}\mu \sum_{i=1}^N \psi_{i,L} \psi_{i,R}\,,
\ee
so that the total Hamiltonian is $H=H_{{\rm SYK},L}+H_{{\rm SYK},R}^*+H_{\mathrm{int}}$. Up to a constant shift, this has the form \eqref{eq:Hamiltonianprop}. We will consider the improved parent Hamiltonian
\be 
H = H_{{\rm SYK},L}+H_{{\rm SYK},R}^*+H_{\mathrm{int}} + \dfrac{\xi}{2N}\left(H_{{\rm SYK},L}-H_{{\rm SYK},R}^*\right)^2\,.
\ee 

Introducing a Hubbard--Stratonovich field $\phi(\tau)$,
\be
\exp\!\left[-\int d\tau\, \frac{\xi}{2N}\big(H_{0,L}-H_{0,R}^{*}\big)^2\right]
=
\int \mathcal{D}\phi\,
\exp\!\left[
-\int d\tau\,
\left(
\frac{N}{2\xi}\phi(\tau)^2
+\mathrm{i}\phi(\tau)\big(H_{0,L}-H_{0,R}^{*}\big)
\right)
\right].
\label{eq:HSphi}
\ee
Thus, the left and right SYK couplings are dressed by
\be
\Lambda_L(\tau)=1+\mathrm{i}\phi(\tau),
\qquad
\Lambda_R(\tau)=1-\mathrm{i}\phi(\tau).
\ee

Consider the bi-local collective fields $G_{ab}(s,s')=\frac{1}{N}\sum_i \psi_i^{\,a}(s)\psi_i^{\,b}(s')$ and $\Sigma_{ab}(s,s')$, where $a,b \in \{ L,R\}$. After disorder averaging, each bilocal interaction kernel picks up a factor $\Lambda_a(\tau_1)\Lambda_b(\tau_2)$, and the Euclidean effective action becomes
\begin{align}
-\frac{I_{E}[G,\Sigma,\phi]}{N}
&=
\log {\rm Pf}\!\big(\delta_{ab}\partial_\tau-\Sigma_{ab}\big)
-\frac12\int d\tau_1 d\tau_2 \sum_{a,b}
\Big[
\Sigma_{ab}G_{ab}-
s_{ab}\frac{\mathcal{J}^2}{2q^2}\,
\Lambda_a(\tau_1)\Lambda_b(\tau_2)\,
G_{ab}^q
\Big]
\nonumber\\
&\hspace{2.3cm}
-\frac{\mathrm{i}\mu}{2}\int d\tau\, G_{LR}(\tau,\tau)
-\int d\tau\,\frac{\phi(\tau)^2}{2\xi},
\label{eq:penalized_collective}
\end{align}
where $s_{LL}=s_{RR}=1$ and $s_{LR}=s_{RL}=(-1)^{q/2}$. The Schwinger–Dyson equations of the model are
\begin{gather}
G_{ab} = (\partial_\tau - \Sigma)^{-1}_{ab}\,,\label{eq:G_penalty_general}\\[.2cm]
\Sigma_{ab}(\tau_1,\tau_2)
=
s_{ab}\frac{\mathcal J^2}{2q}\,
\Lambda_a(\tau_1)\Lambda_b(\tau_2)\,
G_{ab}(\tau_1,\tau_2)^{q-1}
-\mathrm{i}\mu\,\delta_{aL}\delta_{bR}\,\delta(\tau_{12}) \,,
\label{eq:Sigma_penalty_general}\\[.2cm]
\phi(\tau)
=
\frac{\mathrm{i}\mathcal J^2\xi}{4q^2}
\sum_{a,b}s_{ab}\int d\tau'\,
\Big[
\eta_a\,\Lambda_b(\tau')\,G_{ab}(\tau,\tau')^q
+
\eta_b\,\Lambda_a(\tau')\,G_{ab}(\tau',\tau)^q
\Big]\,,
\label{eq:phi_saddle_sym}
\end{gather}
where $\eta_L=+1$ and $\eta_R=-1$.

Setting $\phi(\tau)=0$, the deformed Schwinger--Dyson equations reduce to those of the Maldacena-Qi model \cite{maldaqi}. Moreover, $\phi(\tau)=0$ is itself a solution of \eqref{eq:phi_saddle_sym} because the undeformed saddle satisfies
$G_{LL}(\tau_1,\tau_2)=G_{RR}(\tau_1,\tau_2)$ and the off-diagonal correlator has definite exchange parity,
$G_{LR}(\tau_1,\tau_2)=\pm G_{LR}(\tau_2,\tau_1)$ \cite{maldaqi}.
Therefore, the penalty term does not modify the leading large-$N$ saddle, and its effects first appear through fluctuations around it.

To study these effects, one could, in principle, proceed in several ways: by solving the deformed model numerically, by performing a large-$q$ analysis, or by working in the low-energy Schwarzian regime. The large-$q$ expansion is not very useful for our purposes, since the relevant effects are not calculable at $O(1/q^2)$ \cite{maldaqi}. In what follows, we therefore focus on the low-energy analysis.

\subsection{Low-energy gravitational analysis}

\begin{figure}[h]
    \centering
    \includegraphics[width=0.6\linewidth]{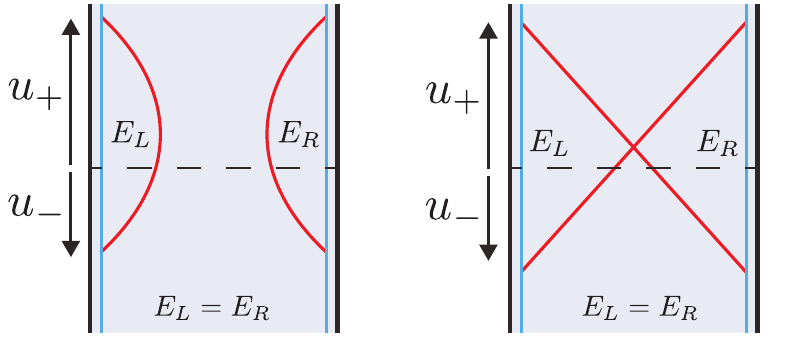}
    \caption{Diagrams contributing to the fidelity loss at $O(\eta^2)$ in the Maldacena-Qi wormhole. These contributions are suppressed by the penalty term because the intermediate configuration generally lies outside the diagonal subspace, with $E_L \neq E_R$ over part of the Euclidean evolution. More precisely, one should integrate over $E_L$ and $E_R$ with the appropriate kernels. Physically, these diagrams describe particle creation after quench, which takes the state out of the diagonal subspace.}
    \label{fig:JTdiagrams}
\end{figure}

Let us first consider $\xi=0$. In the low-energy approximation of \cite{maldaqi}, the $LR$ coupling only modifies the dynamics of the reparametrization modes through a modified effective potential for the gauge-invariant geodesic length variable. At this level of approximation, the fidelity
\be
\mathcal{F} =
\frac{|\langle {\rm TFD}_{\beta_\star}|{\rm GS}\rangle|^2}{\langle {\rm TFD}_{\beta_\star}|{\rm TFD}_{\beta_\star}\rangle \langle {\rm GS}|{\rm GS}\rangle}
\ee
is equal to one. We write the norms explicitly because the states are naturally prepared by Euclidean path integrals. Fig.~\ref{fig:JTdiagrams} shows the nonlocal quench diagrams relevant for the leading correction to $\langle {\rm GS}|{\rm GS}\rangle$.

Moreover, the optimal inverse temperature is \cite{maldaqi}
\be 
\beta_\star = \dfrac{2\pi \alpha_S}{\mathcal{J}} (\eta \Delta_{\mathcal O})^{\frac{1}{2(\Delta_{\mathcal O}-1)}}\,,\qquad 
\eta = \dfrac{\mu \alpha_S}{\mathcal{J}} \dfrac{c_{\Delta_{\mathcal O}}}{(2\alpha_S)^{2\Delta_{\mathcal O}}}\,.
\ee
Throughout this appendix, $\Delta_{\mathcal O}$ denotes the IR conformal dimension of the operator appearing in the interaction term. For the Maldacena-Qi coupling \eqref{eq:intMQ}, this operator is a single fermion, so
\be
\Delta_{\mathcal O}=\Delta_\psi=\frac{1}{q}\,.
\ee
This should not be confused with the gap of the parent Hamiltonian used elsewhere in the main text. In addition, $\alpha_S$ and $c_{\Delta_{\mathcal O}}$ are constants defined in \cite{Maldacena:2016hyu}. Consistency of the low-energy analysis requires
\be
1 \ll \mathcal{J}\beta_\star \ll N\,,
\ee
so the $LR$ coupling strength $\eta$ must be small, but not parametrically too small.

The first correction to the unit fidelity arises from additional diagrams associated with particle creation when the interaction is quenched at Euclidean time $u=0$. More precisely, the leading correction to $\log\mathcal{F}$ is controlled by
\[
\langle O^i_L(u_+)O^i_R(u_+)O^i_L(u_-)O^i_R(u_-)\rangle
-
\langle O^i_L(u_+)O^i_R(u_+)\rangle
\langle O^i_L(u_-)O^i_R(u_-)\rangle
\]
with $u_+>0$ and $u_-<0$. At large $N$, this four-point function factorizes into two Wick contractions: one built from same-side correlators and one built from left-right correlators. On the wormhole background, these give, respectively, the functions
$t'^{4\Delta_{\mathcal O}}\sinh^{-4\Delta_{\mathcal O}}\!\big(\tfrac{t'(u_+-u_-)}{2}\big)$
and
$t'^{4\Delta_{\mathcal O}}\cosh^{-4\Delta_{\mathcal O}}\!\big(\tfrac{t'(u_+-u_-)}{2}\big)$,
for constant $(t')^{2(1-\Delta_{\mathcal O})} = \eta \Delta_{\mathcal O}$ in the wormhole solution. The relative sign between them is fixed by statistics: it is positive for bosonic operators and negative for fermionic ones, as in the present case \eqref{eq:intMQ}, since the crossed contraction requires exchanging the two operators that connect the left and right systems. This leads to \cite{maldaqi}
\be
\log \mathcal{F}
=
-N\eta^2 t'^{4\Delta_{\mathcal O}}
\int_0^\infty {\rm d}u_+ \int_{-\infty}^0 {\rm d}u_- \,
\left[
\pm \frac{1}{\cosh^{4\Delta_{\mathcal O}}\!\left(\frac{t'(u_+-u_-)}{2}\right)}
+
\frac{1}{\sinh^{4\Delta_{\mathcal O}}\!\left(\frac{t'(u_+-u_-)}{2}\right)}
\right].
\label{eq:MQ_overlap_nonlocal_short}
\ee
After the change of variables $x=\tfrac{t'}{2}(u_+-u_-)$, this can be written as
\be 
\log \mathcal{F}
=
-N\eta^2  t'^{4\Delta_{\mathcal O}-2} h_\pm(\Delta_{\mathcal O})\,,
\qquad
h_\pm(\Delta_{\mathcal O})
=
4\int_0^\infty {\rm d}x\,x\,[\pm \cosh^{-4\Delta_{\mathcal O}}x+\sinh^{-4\Delta_{\mathcal O}}x]\,.
\ee 
Thus the fidelity decays exponentially in $N$, albeit with a very small coefficient for operators with $\Delta_{\mathcal O}<1/2$.

In the fermionic case relevant for \eqref{eq:intMQ}, one has $\Delta_{\mathcal O}=\Delta_\psi=1/q$. The large-$q$ expansion then gives
\be
h_-(\Delta_\psi)\approx 7\zeta(3)\Delta_\psi\,,
\ee
while $\eta^2 t'^{4\Delta_\psi-2}=t'^2/\Delta_\psi^2$. Therefore
\be
\log \mathcal F\approx -7\zeta(3)N\, \frac{t'^2}{\Delta_\psi}\,.
\ee
Using $t'=2\pi \alpha_S/(\mathcal J \beta_\star)$ together with $\alpha_S\approx 1/(4q^2)$, we obtain
\be
\log \mathcal F\approx 
-\,
\frac{21 N}{q^3(\mathcal J\beta_\star)^2}
\,.
\ee
Thus, when $\mathcal J\beta_\star$ is held fixed in the large-$q$ limit, the fidelity-loss exponent scales as $1/q^3$. This behavior relies on scaling the $LR$ coupling as $\mu=\hat\mu/q$ with fixed $\hat{\mu}$. If instead $\mu$ is kept fixed, the suppression is weaker, scaling only as $1/q^2$. In either case, for $\mathcal{J}\beta_\star \gg 1$ the coefficient is numerically small, even at $q=4$, which may help explain why this loss of fidelity is difficult to detect numerically at small system sizes \cite{maldaqi,Garcia-Garcia:2019poj,Lantagne-Hurtubise:2019svg,Alet:2020ehp,Su:2020zgc}. For bosonic operators there is no large-$q$ suppression, since there is no relative sign between the direct and crossed contractions, and
\be
h_+(\Delta_{\mathcal O}) \approx \frac{1}{2\Delta_{\mathcal O}^2}
\ee
at small $\Delta_{\mathcal O}$.

We now explain how this is modified by the penalty term when $\xi \neq 0$ (see \cite{Magan:2024aet,Magan:2025hce} where this term was considered in the gravity context). At low energies, the deformation is naturally represented as a penalty on the relative Schwarzian energy,
\be
\Delta S_{\rm pen}
=
\frac{\mathcal{J}\xi}{2N \alpha_S}\int {\rm d}u\,\big(E_L(u)-E_R(u)\big)^2.
\label{eq:S_pen_low_short}
\ee
This term does not affect the leading saddle point, but it does modify the fluctuations. We do not, however, expect it to change the coupled-Schwarzian fluctuation analysis of \cite{maldaqi}, since gauge invariance sets the boost energy to zero. For instance, we expect the gap to be unchanged. By contrast, the diagrams in Fig.~\ref{fig:JTdiagrams} contributing to \eqref{eq:MQ_overlap_nonlocal_short} go beyond this approximation and are modified, being reweighted by a factor $e^{-\Delta S_{\rm pen}}$. Formally, this means that one should insert this extra factor into the path integral for the quench diagrams.

To obtain a simple estimate, we approximate the intermediate configuration created by the nonlocal diagram as carrying an approximately constant non-extensive boost energy
\be
E_*/\mathcal J \sim \Delta_{\mathcal O}
\ee
during the Euclidean interval $|u_+-u_-|$. A more refined treatment would evaluate these diagrams using the technology developed in, for example, \cite{Mertens:2017mtv}, suitably adapted to the present setup, where one integrates over left and right energies $E_L$ and $E_R$ against appropriate kernels. Within our simple approximation, the penalty term contributes the factor
\[
\exp\!\left[-\frac{\mathcal{J}\xi}{2N\alpha_S}|u_+-u_-|E_*^2\right],
\]
so the diagrams are damped at large separations compared to
\be 
\alpha = \frac{ \xi\beta_\star  E_*^2}{2\pi N}\,.
\ee 
The overlap correction is then obtained from \eqref{eq:MQ_overlap_nonlocal_short} by multiplying the integrand by this factor. Equivalently, after changing variables to $x=\tfrac{t'}{2}(u_+-u_-)$, one replaces $h_\pm(\Delta_{\mathcal O})$ by
\[
h_\pm^{(\xi)}(\Delta_{\mathcal O})
=
4\int_0^\infty {\rm d}x\,x\,e^{-\alpha x}
\left[\pm \cosh^{-4\Delta_{\mathcal O}}x+\sinh^{-4\Delta_{\mathcal O}}x\right].
\]
For small $\alpha$, this gives a perturbative reduction of the original answer. For large $\alpha \gg 1$, the integral is dominated by small $x$, and one finds
\be 
h_\pm^{(\xi)}(\Delta_{\mathcal O})
\sim
4\Gamma(2-4\Delta_{\mathcal O})\alpha^{-(2-4\Delta_{\mathcal O})}
\ee
for $\Delta_{\mathcal O}<1/2$. In this regime the nonlocal correction is strongly suppressed, leaving only a residual contribution from very short Euclidean separations near the quench:
\be
\log \mathcal{F}
=
-N\eta^2  t'^{4\Delta_{\mathcal O}-2}\,4\Gamma(2-4\Delta_{\mathcal O})\alpha^{-(2-4\Delta_{\mathcal O})}\,,
\qquad
\mathcal{J}\xi \gg \dfrac{ N}{\mathcal{J}\beta_\star}\,.
\ee
Thus the crossover between exponential fidelity decay and saturation occurs when
\be 
\mathcal{J}\xi \sim N^{\frac{3-4\Delta_{\mathcal O}}{2-4\Delta_{\mathcal O}}}\,.
\ee 
For the $q=4$ fermionic interaction \eqref{eq:intMQ}, one has $\Delta_{\mathcal O}=\Delta_\psi=1/4$, and this gives
\be
\mathcal J \xi \sim N^2\,.
\ee
This should be viewed as a simple effective estimate of how the penalty term improves the fidelity by suppressing off-diagonal leakage in the Maldacena-Qi wormhole. We expect the precise crossover scale to depend on the system, as suggested for example by the numerics in the mixed-field Ising model shown in Fig.~\ref{fig:regimes}. More generally, this fits naturally with the estimate in \eqref{eq:offdiagweightpenalty}: in the Maldacena-Qi wormhole, the low-frequency off-diagonal spectral weight depends on $\Delta_{\mathcal O}$, and that changes the power of $N$ needed for the penalty term to become effective.

\end{document}